\documentclass[pra,showpacs,twocolumn]{revtex4}
\usepackage{graphicx}
\usepackage{amssymb}
\usepackage{amscd}
\usepackage{amsmath}
\newcommand{\nn}{\nonumber}
\newcommand{\rd}{{\rm d}}
\newcommand{\re}{{\rm e}}
\newcommand{\ri}{{\rm i}}
\newcommand{\bbs}{{\bf s}}
\newcommand{\kx}{\ensuremath{\hat s_x}}
\newcommand{\ky}{\ensuremath{\hat s_y}}
\newcommand{\kz}{\ensuremath{\hat s_z}}
\newcommand{\kkx}{\ensuremath{s_x}}
\newcommand{\kky}{\ensuremath{s_y}}
\newcommand{\kkz}{\ensuremath{s_z}}
\newcommand{\fmn}{\ensuremath{\hat F(\kz)}}
\newcommand{\kfmn}{\ensuremath{f(\kkz)}}

\newcommand{\gmn}{\ensuremath{\hat G(\kz)}}
\newcommand{\kgmn}{\ensuremath{g(\kkz)}}
\newcommand{\ah}{\ensuremath{\hat a}}
\newcommand{\ad}{\ensuremath{\hat a^\dagger}}
\newcommand{\am}{\ensuremath{\hat a^{\dagger m}}}
\newcommand{\aj}{\ensuremath{\hat a^{\dagger j}}}
\newcommand{\bh}{\ensuremath{\hat b}}
\newcommand{\bd}{\ensuremath{\hat b^\dagger}}
\newcommand{\bn}{\ensuremath{\hat b^{\dagger n}}}
\newcommand{\hn}{\ensuremath{\hat N}}

\newcommand{\kah}{\ensuremath{a}}
\newcommand{\kad}{\ensuremath{a^*}}
\newcommand{\kam}{\ensuremath a^{* m}}
\newcommand{\kamm}{\ensuremath a^{* m-1}}
\newcommand{\kbh}{\ensuremath{b}}
\newcommand{\kbd}{\ensuremath{b^*}}
\newcommand{\kbn}{\ensuremath{b^{* n}}}
\newcommand{\pmn}{\ensuremath{\hat P(\kz)}}
\newcommand{\pmnm}{\ensuremath{\hat P (\kz\!\!-\! 1)}}
\newcommand{\joh}{\ensuremath{\hat J_0}}
\newcommand{\jod}{\ensuremath{\hat J_0^\dagger}}
\newcommand{\jp}{\ensuremath{\hat J_+}}

\newcommand{\jm}{\ensuremath{\hat J_-}}
\newcommand{\jpm}{\ensuremath{\hat J_\pm}}
\newcommand{\jmd}{\ensuremath{\hat J_-^\dagger}}
\newcommand{\jz}{\ensuremath{\hat J_z}}

\newcommand{\jx}{\ensuremath{\hat J_x}}

\newcommand{\jy}{\ensuremath{\hat J_y}}

\begin{document}
\title[Classical-quantum correspondence in bosonic two-mode conversion systems]{Classical-quantum correspondence in bosonic two-mode conversion systems: polynomial algebras and Kummer shapes}
\author{Eva-Maria Graefe$^{1}$}
\author{Hans J\"urgen Korsch$^2$}
\email{korsch@physik.uni-kl.de}
\author{Alexander Rush$^{1}$}
\affiliation{${}^1$ Department of Mathematics, Imperial College London, London, SW7 2AZ, United Kingdom\\
${}^2$ FB Physik, TU Kaiserslautern, D--67653 Kaiserslautern, Germany}
\date{\today}

\begin{abstract} 
Bosonic quantum conversion systems can be modeled by many-particle single-mode
Hamiltonians  describing a conversion of $n$ molecules of type A into $m$ molecules of type B and vice versa. These Hamiltonians are analyzed in terms of generators of a 
polynomially deformed $su(2)$ algebra. In the mean-field
limit of large particle numbers, these systems become classical and their Hamiltonian dynamics can  again be described by polynomial
deformations of a Lie algebra, where quantum commutators are replaced by 
Poisson brackets. The Casimir operator restricts the motion to Kummer shapes, 
deformed Bloch spheres with cusp singularities depending on $m$ and $n$. It is demonstrated that the many-particle eigenvalues can be recovered from the mean-field dynamics using a WKB type quantization condition. The many-particle state densities can be semiclassically approximated by
the time-periods of periodic orbits, which show characteristic steps and
singularities related to the fixed points, whose bifurcation properties are analyzed.
\end{abstract}


\maketitle

\section{Introduction}
In a recent paper \cite{Grae15} some of the authors  studied bosonic atom-molecule conversion systems describing (non-interacting) atoms which  can undergo a conversion
to diatomic molecules, both populating a single mode. This is the simplest possible conversion system modeling atom diatomic molecule conversion in cold atom systems
and Bose-Einstein condensates (BECs). These systems have been studied extensively  \cite{Kara94,Vard01,Kara02,Zhou03,Sant06,LiYe09,Liu10,Khri11,LiFu11,Pere11,Sant11,Cui12,Shen13, Donl02},
quite often in a mean-field approximation \cite{Vard01,Sant06,LiYe09,Pere11,Grae15}
where the conversion can be described in terms of classical dynamics.
In addition, the influence of particle interaction \cite{Sant06,LiYe09,LiFu11,Cui12},
noise \cite{Shen13} and particle losses \cite{Cui12} has been studied as well as
extensions to systems coupling two modes \cite{Rela14}.

The  mean-field approximation of the many-particle system 
derived in \cite{Grae15} based on a polynomially
deformed $su(2)$ algebra \cite{Roce90,Poly90,Bona96,Lee10} showed
that the mean-field conversion dynamics takes place on a deformed
Bloch sphere of a teardrop shape (see also \cite{Cui12}). Such surfaces 
also appear in the different context of classical harmonic
oscillators at a $1\!:\!2$ resonance. In this context these surfaces have been
denoted as Kummer shapes \cite{Holm11a,Holm12}, named after
preceding work by Kummer \cite{Kumm76,Kumm78,Kumm81,Kumm86,Kumm90}.

Here we extend the work in \cite{Grae15} to more general conversion systems, 
where $m$ molecules of type A can form $n$ molecules of type B and vice versa, 
conserving the total number of particles.
The corresponding Hamiltonian discussed in the subsequent section models, e.g., 
polyatomic homonuclear molecular BECs \cite{Sant11,Dou12, Dou13}, however,
in addition to these applications in cold atom physics, it also describes other
systems of interest in different areas of physics, as for example higher order
harmonic generation, multiphoton processes,
frequency conversion or, quite generally, the superposition 
of two harmonic oscillators. 

For such systems, nonlinear polynomial algebras \cite{Roce90,Poly90,Delb93} arise
in a natural way  \cite{Debe98,Debe00,Beck00,Liu10,Lee10,Kara94,Kara02,Link03}. 
However they have been almost exclusively 
employed in context with superintegrability (or supersymmetry)
\cite{Debe98,Debe00,Beck00} allowing
an analytic evaluation of the energy spectrum by means of an algebraic
Bethe ansatz \cite{Link03,Zhou03,Sant11}. Here we employ this
algebraic approach to demonstrate an interesting connection
between these quantum nonlinear algebras to corresponding
ones in classical mechanics where quantum commutators are replaced by
Poisson brackets in a mean-field approximation for large $N$.
Here the general $n\!:\!m$-Kummer shapes \cite{Holm11a,Holm12}
replace the familiar Bloch sphere of the $1\!:\!1$ case.
 
Algebraic methods are employed in most studies of  many-particle conversion 
models, as for example  the combined Heisenberg-Weyl and $su(1,1)$
algebras in \cite{Pere11} for atom-diatom conversion. Here we will
employ polynomially deformed algebras appearing in a Jordan-Schwinger transformation,
which is described in the following section. 
The corresponding mean-field system is derived in the subsequent section, 
followed by a numerical comparison between the many particle energies and the 
mean-field energies. We then apply a quantization method to the mean-field system to 
demonstrate how many-particle energies can be acccurately recovered from the classical 
system, before finally comparing the mean-field period and the many-particle density of
states. We end with a summary and an outlook.

\section{Quantum many-particle conversion systems}
\subsection{The Hamiltonian}
A toy model for studying multi-particle conversion systems is provided by 
the Hamiltonian 
\begin{equation}
\label{hamil-1}
\hat H_0=\epsilon_a\hat a^\dagger \hat a+\epsilon_b\hat b^\dagger \hat b
+\tfrac{v}{2\sqrt{N^{m+n-2}}}\,\big(\hat a^{\dagger\,m} \hat b^n+\hat a^m\hat b^{\dagger\,n}\big),
\end{equation}
where $\hat a^\dagger$, $\hat a$ and $\hat b^\dagger$, $\hat b$
with $[\ah,\ad]=[\bh,\bd]=1$,  $[\ah,\bh]=[\ah,\bd]=0$ are the molecular creation and
annihilation operators of molecules of type A or type B, respectively, and
$\epsilon_{a,b}$ are the energies of the molecular modes. 
The conversion of $m$ molecules A into $n$ molecules B and vice versa conserves
the total number $N$ of particles, i.e., 
\begin{equation}
\label{atomnumber}
\hat N=n\hat a^\dagger \hat a+m\hat b^\dagger \hat b
\end{equation}
commutes with the Hamiltonian: $\big[\hat H_0,\hat N\big]=0$. In (\ref{hamil-1}).
The parameter $v$ describes the conversion strength per particle, and,
due to the $N$-dependent scaling factor of the conversion strength, all terms in the Hamiltonian scale linearly with $N$. 

For the special cases where either $n=1$ or $m=1$ the Hamiltonian \eqref{hamil-1} describes the association and dissociation of polyatmic molecules \cite{Dou12,Dou13}. For $n=m$, the term $\hat a^{\dagger\,m} \hat b^n$ and its Hermitian conjugate can describe tunneling of groups of atoms between the two states, a common example is pair tunneling \cite{Mazz06,Eckh08}. The trivial case $n=1=m$ reduces to the tunneling of individual bosonic particles in a two-mode system.

As the particle number $N$ is conserved we can drop the constant $\hat N$-dependent term 
$(m\epsilon_a+n\epsilon_b)\tfrac{\hn}{2mn}$ from the Hamiltonian \eqref{hamil-1} and  from now on we consider 
\begin{eqnarray}
\label{hamil-2}
\hat H=\epsilon\frac{n\ad\ah-m\bd\bh}{2mn} +\tfrac{v}{2\sqrt{N^{m+n-2}}}\,\big(\hat a^{\dagger\,m} \hat b^n+\hat a^m\hat b^{\dagger\,n}\big)
\end{eqnarray}
with 
\begin{eqnarray}
\epsilon=m\epsilon_a-n\epsilon_b.
\end{eqnarray}

In the present paper we focus on the effect of the conversion term in the Hamiltonian, and disregard interactions between the particles, which could be included as terms of the form $\hat a^\dagger \hat a\,\hat a^\dagger \hat a$, 
$\hat b^\dagger \hat b\,\hat b^\dagger \hat b$ and 
$\hat a^\dagger \hat a\,\hat b^\dagger \hat b$ \cite{Sant06,LiYe09,LiFu11}. Note that while these interactions would make the classical dynamics and the quantum spectra more complicated, the algebraic approach presented here carries through to the case with interactions. In particular, the Hamiltonian can still be expressed in terms of deformed $SU(2)$ algebras.
\subsection{Quantum polynomial algebras}
\label{s-qu-pol-alg}
The analysis of the system described by the Hamiltonian (\ref{hamil-1}) is greatly simplified
by using techniques recently developed as deformed Lie algebras, more precisely polynomial
deformations of the $su(2)$ algebra (see Appendix \ref{s-poly-alg}). Here we will
closely follow the analysis by Lee {\it et al.} \cite{Lee10}. We first
introduce the generalized Jordan-Schwinger mapping to the operators
\begin{eqnarray}
\label{kxyz}
\kx &=&\frac{ \am\bh^n+\ah^m\bn}{2\sqrt{N^{m+n-2}}}\,,\ 
\ky =\frac{\am\bh^n-\ah^m\bn}{2\ri \sqrt{N^{m+n-2}}}\,,\ \nn\\
\kz& =&\frac{n\ad\ah-m\bd\bh}{2mn},
\end{eqnarray}
which commute with the number operator $\hat N$ in (\ref{atomnumber}).
All these operators scale linearly with the particle number $N$. In analogy to the case of a simple two-mode system, the operator $\kz$ encodes the population imbalance between molecules of type $A$ and type $B$, and the operators $\hat s_{\pm}=\kx\pm\ri\ky$ describe the conversion from $A$ to $B$ and vice versa. 

According to \cite{Lee10}, 
the commutation relations can be written as
\begin{eqnarray}
\label{kxyz-comm-1}
[\kz,\kx]\!=\!\ri\ky,\,[\ky,\kz]\!=\!\ri\kx,\, 
[\kx,\ky]\!=\!\ri\fmn,
\end{eqnarray}
where $\fmn$ is a polynomial of order $n+m$ in $s_z$, and can be expressed as
\begin{equation}
\fmn=-\frac{n^n m^m}{2N^{m+n-2}}\,\big(\pmn-\pmnm\big)\,,\label{fop}
\end{equation}
with (see Appendix \ref{s-poly-alg} for details) 
\begin{eqnarray}
\label{Pmn}
\hat P(\kz)\!=\!
\Pi_{\mu=1}^m\big(\tfrac{\hn}{2mn}\!+\,\kz\!+\!\tfrac{\mu}{m}\big)
\Pi_{\nu=1}^n\big(\tfrac{\hn}{2mn}\!-\!\kz\!-\! 1\!+\!\tfrac{\nu}{n}\big).
\end{eqnarray}
The Casimir operator for the algebra is given by
\begin{eqnarray}
\label{Fop}
\hat C=\kx^2+\ky^2+\gmn,
\end{eqnarray}
with 
\begin{equation}
\gmn=-\frac{n^n m^m}{2N^{m+n-2}}\,\big(\pmn+\pmnm\big).\label{gop}
\end{equation}
Obviously the Casimir operator \eqref{Fop} can be modified by adding terms depending only on the number operator $\hn$, which 
also commutes with the $\hat s_j$.

If $n$ and $m$ are interchanged, $(m,n) \longleftrightarrow  (n,m)$,
the polynomials $\hat F(\kz)$ and $\hat G(\kz)$ transform according to
\begin{eqnarray}
\label{fgsym1_m}
\hat F(\kz) \longleftrightarrow -\hat F(-\kz)\  ,\ 
\hat G(\kz) \longleftrightarrow \hat G(-\kz),
\end{eqnarray}
and thus for $m=n$ they have the symmetries
\begin{eqnarray}
\label{fgsym2_m}
\hat F(-\kz)=-\hat F(\kz) \quad \textrm{and}\quad  
\hat G(-\kz)=\hat G(\kz)\,,
\end{eqnarray}
i.e.~$\hat F(\kz)$ and $\hat G(\kz)$ are odd or even polynomials. 

In terms of the operators (\ref{kxyz}) the Hamiltonian (\ref{hamil-2}) can be rewritten as
\begin{eqnarray}
\label{hamil-3}
\hat H=\epsilon \kz+v\kx,
\end{eqnarray}
which is the Hamiltonian referred to in the following.

The Heisenberg equations of motion $\ri \dot {\hat A}=[\hat A,\hat H]$
for the operators  (\ref{kxyz}) read
\begin{eqnarray}
\label{heis-k}
\frac{\rd}{\rd t} \kx&=&-\epsilon \ky,\nonumber\\
\frac{\rd}{\rd t} \ky&=&\epsilon \kx-v\fmn,\\
\frac{\rd }{\rd t}\kz&=&v \ky,
\nonumber
\end{eqnarray}
which conserve, in addition to the particle number $\hat N$, the Casimir operator
$\hat C(\kx,\ky,\kz)$, i.e.
\begin{eqnarray}
\label{Kx2Ky2}
\kx^2+\ky^2=\hat C-\gmn,
\end{eqnarray}
and therefore
$\langle\kx^2\rangle+\langle\ky^2\rangle=\langle\hat C\rangle-\langle\gmn\rangle$,
corresponding to a generalized Bloch sphere \cite{Shen13}, i.e.~a deformation 
of the Bloch sphere also denoted as a quantum Kummer shape \cite{Odzi14}
in view of the classical Kummer shapes discussed
in the mean-field approximation in section \ref{s-clas-sys}.

Let us discuss some cases considered in the following section in more detail, where in view of the symmetry (\ref{fgsym1_m}) it is sufficient to 
study the cases $m\ge n$. \\[2mm]
\noindent
(1) $(m,n)=(1,1)$\,:\,  In this linear case of a simple $N$ particle two-mode system we encounter the $su(2)$ algebra, with\begin{eqnarray}
\label{fg11}
\fmn=\kz\ ,\quad \gmn=\kz^2-\left(\tfrac{\hn}{2mn}\right)^2-\tfrac{\hn}{2mn}
\end{eqnarray}
and the Casimir operator 
\begin{eqnarray}
\label{c11}
\hat C=\kx^2+\ky^2+\kz^2-\left(\tfrac{\hn}{2mn}\right)^2-\tfrac{\hn}{2mn}\,. 
\end{eqnarray}
Up to insignificant $\tfrac{\hn}{2mn}$-dependent terms this operator 
is known as $\hat L^2$ for the angular momentum algebra.
The Casimir operator then imposes a restriction to the surface of the
Bloch sphere
\begin{eqnarray}
\label{qKu11}
\kx^2+\ky^2=
\hat C+\left(\tfrac{\hn}{2mn}\right)^2+\tfrac{\hn}{2mn}-\kz^2\,. 
\end{eqnarray}

\noindent
(2) $(m,n)=(2,1)$\,:\, In this case, describing a conversion of two atoms  into  
diatomic molecules, which has been studied quite extensively (see \cite{Grae15}
and references therein), we have
\begin{eqnarray}
\label{fg21-1}
&&\fmn
\!=\!\tfrac{6}{N}\kz^2+\tfrac{\hn}{N}\kz-\tfrac{\hn^2}{8N}
-\tfrac{\hn}{2N},\\
&&\gmn
\!=\!\tfrac{4}{N}\kz^3\!+\!\tfrac{\hn}{N}\kz^2\!+\!\tfrac{8-\hn^2-4\hn}{4N}\kz
\!-\!\tfrac{4\hn^3}{N}\!+\!\tfrac{4\hn^2}{N},
\end{eqnarray}
in agreement with \cite{Grae15} up to the $\kz$-independent terms.\\[2mm]
\noindent
(3) $(m,n)=(2,2)$\,:\,    For this case the nonlinear algebra corresponds to
the (cubic) Higgs algebra (see \cite{Debe98,Beck00} and references therein) with
\begin{eqnarray}
\label{fg21-2}
\fmn
&\!\!=\!\!&\tfrac{4}{N^2}\!\left(\!-8\kz^3+\left(8\left(\!\tfrac{\hn}{2mn}\!\right)^2+4\tfrac{\hn}{2mn}-1\!\right)\!\kz\!\right) \\
\gmn&\!\!=\!\!&\tfrac{4}{N^2}\!\left(\!-4\kz^4+\left(8\left(\!\tfrac{\hn}{2mn}\!\right)^2\!\!+4\tfrac{\hn}{2mn}-5\right)\kz^2\right.\nn\\
&&\left.\quad -4\!\left(\!\tfrac{\hn}{2mn}\!\right)^4\!\!\!-\!4\!\left(\!\tfrac{\hn}{2mn}\!\right)^3\!\!\!+\!\left(\!\tfrac{\hn}{2mn}\!\right)^2\!\!\!+\!\tfrac{\hn}{2mn}\right).
\end{eqnarray}
Note that $\fmn$ is an odd polynomial in $\kz$ and $\gmn$ is even, as expected from (\ref{fgsym2_m}).

In the same way the cases  $m,n\ge 3$ can be written as explicit polynomials, if desired.  The coefficients of the polynomials $F$ and $G$ can in general be related as discussed in Appendix \ref{s-poly-alg}.
\subsection{Matrix representation for numerical calculations}
\label{s-fock}
The dimension of the Hilbert space is $\left[\frac{N}{mn}\right]_<+1$, in what follows we shall assume that $N$ is
an integer multiple of $mn$. As a consequence of the superintegrability the eigenvalues of the Hamiltonians
can be obtained analytically \cite{Sant11,Lee10} by means of a Bethe 
ansatz \cite{Link03}, but for the following results we 
simply diagonalize $\hat H$  numerically using a Fock basis 
\begin{eqnarray}
\label{fock-1}
|j,k\rangle=\tfrac{1}{\sqrt{j!\,k!}}\,\aj\hat b^{\dagger k}\,|0,0\rangle\ ,\quad 
j,k=0,\,1,\,\ldots,
\end{eqnarray}
where $j$ and $k$ denote the numbers of molecules of type $A$ and $B$ respectively. The states $|j,k\rangle$ form an orthonormal basis of the Hilbert space, that is, $\langle j',\, k'|j,\, k\rangle=\delta_{j',j}\delta_{k',k}$. We have 
\begin{eqnarray}
\langle j',\, k'|\ad\ah| j,\, k\rangle&=&j\,\delta_{j',j}\delta_{k',k},\nn\\
\langle j'\, k'|\am| j\, k\rangle&=&{\textstyle \sqrt{\frac{(j+m)!}{j!}}}\,\delta_{j',j+m}\delta_{k',k},\\
\langle j',\, k'|\ah^m| j,\, k\rangle&=&{\textstyle \sqrt{\frac{j!}{(j-m)!}}}\,\delta_{j',j-r}\delta_{k',k}\nn,
\end{eqnarray}
and similarly for $\bh$ with $j$ replaced by $k$. The $\left(\frac{N}{mn}+1\right)$-dimensional subspaces of eigenstates of $\hn$
with eigenvalue $N$ are spanned by the basis
\begin{eqnarray}
\label{fock-2}
|\mu\rangle=|\mu m,\tfrac{N}{m}-\mu n\rangle \ ,\quad \mu=0,\,1,\ldots,\,\tfrac{N}{mn}.
\end{eqnarray}
Then the operators $\kx$, $\ky$, $\kz$ are represented by the matrices
\begin{eqnarray}
\label{k-mat-1}
\langle\mu'|\kx|\mu\rangle&=&
\tfrac{1}{2}\big(\sqrt{\beta_{\mu+1}}\,\delta_{\mu',\mu+1}
+\sqrt{\beta_\mu}\,\delta_{\mu',\mu-1}\big)\,,\\[2mm]
\label{k-mat-2}
\langle\mu'|\ky|\mu\rangle&=&
\tfrac{1}{2\ri }\big(\sqrt{\beta_{\mu+1}}\,\delta_{\mu',\mu+1}-\sqrt{\beta_\mu}\,\delta_{\mu',\mu-1}\big)\,,\\[2mm]
\label{k-mat-3}
\langle\mu'|\kz|\mu\rangle&=&\big(\mu-\tfrac{N}{2mn}\big)
\delta_{\mu',\mu}\,.
\end{eqnarray}
with  
\begin{equation}
\beta_\mu=\frac{1}{N^{m+n-2}}\frac{(\mu m)!}{\left(\mu m-m\right)!}\frac{(\frac{N}{m}-\mu n+n)!}{(\frac{N}{m}-\mu n)!}.
\end{equation}
The matrices representing $\kx$  and $\ky$ are tridiagonal
and the matrix $\kz$ is diagonal with equidistant eigenvalues
ranging from $-\tfrac{N}{2mn}$ for $\mu=0$ to $+\tfrac{N}{2mn}$ for
$\mu=\tfrac{N}{mn}$. Trivially $\hn$ is equal to the identity multiplied by $N$, so
that also $\fmn$ and  $\gmn$ are diagonal. 
\section{Classical mean-field systems}
\label{s-clas-sys}
\subsection{The mean-field limit}
In the mean-field limit $N\rightarrow \infty$, also denoted as thermodynamic limit,  the quantum operators $\hat A(\ah,\ad)$ are
replaced by $c$ functions $A(\kah,\kad)$  and the quantum commutator 
$[\hat A,\hat B]$ by the Poisson bracket 
$\ri\,\{A,B\}=\partial_\kah A\partial_{\kad} B-\partial_\kah B \partial_{\kad} A
+\partial_\kbh A\partial_{\kbd} B-\partial_\kbh B \partial_{\kbd} A$. 
In order to derive this thermodynamic limit for the systems discussed above, 
we follow two different routes:\\[2mm]
(a) First, as in \cite{Grae15}, we consider the limit of large Hilbert space dimension, similar to a classical limit, with the small parameter
$\eta=( \frac{N}{mn}+1)^{-1}\rightarrow 0$, where only the leading order terms
of the algebra survive.  With the replacement $\eta \hat A\rightarrow  A$ and
$\eta^2 [\hat A,\hat B]\rightarrow  \ri\,\{A,B\}$ the commutator relations
(\ref{kxyz-comm-1}) transform to
\begin{eqnarray}
\label{kxyz-comm-2}
\big\{\kkz,\kkx\big\}=\kky\ ,\ \big\{\kky,\kkz\big\}=\kkx\ ,\ 
\big\{\kkx,\kky\big\}=\kfmn.
\end{eqnarray}
Where $f(\kkz)$ is deduced from the identification $\eta\,\fmn \rightarrow \kfmn\nn$ in leading order of $\eta$ as
\begin{eqnarray}
\kfmn&=& \tfrac{1}{2}m^{2-n}n^{2-m}\,\Big(n\,\big(\tfrac12+\kkz\big)^m\big(\tfrac12-\kkz\big)^{n-1}\nn\\
&&\qquad\qquad  -m\big(\tfrac12+\kkz\big)^{m-1}\big(\tfrac12-\kkz\big)^n\Big),\label{kFasy}
\end{eqnarray}
where we have used the identification $\eta \hn\rightarrow mn$.
The Casimir operator (\ref{Fop}) translates via $\eta ^2 \hat C \rightarrow C$ to         
\begin{eqnarray}
\label{kummer}
C(\kkx,\kky,\kkz)=\kkx^2+\kky^2+g(\kkz),
\end{eqnarray}
where with $ \eta^2\,\gmn \rightarrow \kgmn$  we have 
\begin{equation}
\kgmn= -m^{2-n}n^{2-m}\,\big(\tfrac12+\kkz\big)^m\big(\tfrac12-\kkz\big)^{n}\,.\label{kGasy}
\end{equation}
For the special case $(m,n)=(2,1)$ this yields
\begin{eqnarray}
\label{kFGasy}
f(\kkz)
&=&-\tfrac{1}{4}+\kkz+3\kkz^2\\
g(\kkz)&=&-\tfrac{1}{4}-\tfrac{1}{4}\kkz+\kkz^2+2\kkz^3
\end{eqnarray}
in agreement with  \cite{Grae15}.\\[2mm]
\noindent
(b)  Alternatively, one can start from the classicalized version of the
Hamiltonian (\ref{hamil-1}),
\begin{equation}
\label{hamil-cl}
H_0=\epsilon_a a^*a+\epsilon_bb^*b
+v'\big(a^{*\,m}b^n+a^mb^{*\,n}\big),
\end{equation}
where the operators  $\ah$, $\ad$  are replaced by c-numbers  $\kah$, $\kad$,
i.e. $\eta\,\ah \ad \rightarrow \kad\kah $
with $\{\kah,\kad\}=-\ri$. The equations of motion $\dot A=\{A,H_0\}$
conserve the function $n\kad\kah+m\kbd\kbh$, whose value is the
limit $\eta (n\ad\ah+m\bd\bh)=\eta \hn\rightarrow mn$.
In analogy to  (\ref{kxyz}) we define
\begin{eqnarray}
\kkx &=&\frac{ \kam\kbh^n+\kah^m\kbn}{2\sqrt{(nm)^{m+n-2}}},\ 
\kky =\frac{\kam\kbh^n-\kah^m\kbn}{2\ri \sqrt{(nm)^{m+n-2}}},\nn\\
\kkz&=&\frac{n\kad\kah-m\kbd\kbh}{2mn}.\label{kxyz-cl}
\end{eqnarray}
with 
\begin{equation}
\label{a*ab*b}
a^*a=m\big(\tfrac12+\kkz\big)\quad \textrm{and} \quad b^*b=n\big(\tfrac12-\kkz\big).
\end{equation}

The  Poisson bracket
relations and the Casimir function can be easily evaluated using
$\{\kah,\kam\}=-\ri m \kamm$ as well as
\begin{equation}
\label{an-adn-pois}
\big\{a^m,a^{*\,m}\big\}=-\ri m^2 (a^*a)^{m-1}
\end{equation}
in agreement with the leading order term of the quantum 
commutator given in the Appendix in equation (\ref{an-adn-comm-lo}), and can be
found in Holm's book {\it Geometric Mechanics\/} \cite{Holm11a}. The results are
again the Poisson brackets (\ref{kxyz-comm-2}) 
with the same function $\kfmn$ obtained in (\ref{kFasy}) and finally from
(\ref{kxyz-cl}) and (\ref{a*ab*b})
\begin{eqnarray}
\kkx^2+\kky^2&=&(mn)^{2-m-n}\,(a^*a)^m(b^*b)^n\\
&=&m^{2-n}n^{2-m}\big(\tfrac12+\kkz\big)^m\big(\tfrac12-\kkz\big)^n
=-g(\kkz),\nn
\end{eqnarray}
which defines the functional relation  $C(\kkx,\kky,\kkz)=\kkx^2+\kky^2+g(\kkz)$ 
\cite{Holm11a},
exactly as obtained above.
\subsection{Classical polynomial algebras}
\label{s-pol-alg}
We therefore obtain the Poisson bracket relations
\begin{eqnarray}
\label{kxyz-comm-3}
\big\{\kkz,\kkx\big\}=\kky\ ,\ \big\{\kky,\kkz\big\}=\kkx\ ,\ 
\big\{\kkx,\kky\big\}=\kfmn,
\end{eqnarray}
and we define the function
\begin{eqnarray}
\label{kcasi-2}
C(\kkx,\kky,\kkz)=\kkx^2+\kky^2+\kgmn\,,
\end{eqnarray}
where $f(\kkz)$ and $g(\kkz)$ are given in (\ref{kFasy}) and (\ref{kGasy}).
One can easily check that these functions satisfy 
\begin{eqnarray}
\label{g-f-cl}
\frac{\rd \kgmn}{\rd \kkz}=2\,\kfmn
\end{eqnarray}
 and therefore, using $\big\{\kkx,h(\kkz)\big\}=-\kky h'(\kkz)$ and
  $\big\{\kky,h(\kkz)\big\}=\kkx h'(\kkz)$, we find the relations
\begin{eqnarray}
\label{cas-0}
\big\{C,\kkx\big\}=\big\{C,\kky\big\}=\big\{C,\kkz\big\}=0\,.
\end{eqnarray}
This is a polynomial deformation of the Lie algebra with a Poisson bracket instead
of a commutator and the Casimir function $C(\kkx,\kky,\kkz)$, i.e.~it is a constant of motion for
Hamiltonians $H(\kkx,\kky,\kkz)$. 
\subsection{Dynamics on Kummer shapes}
\label{s-dyn}
\begin{figure}
\begin{center}
\includegraphics[width=30mm]{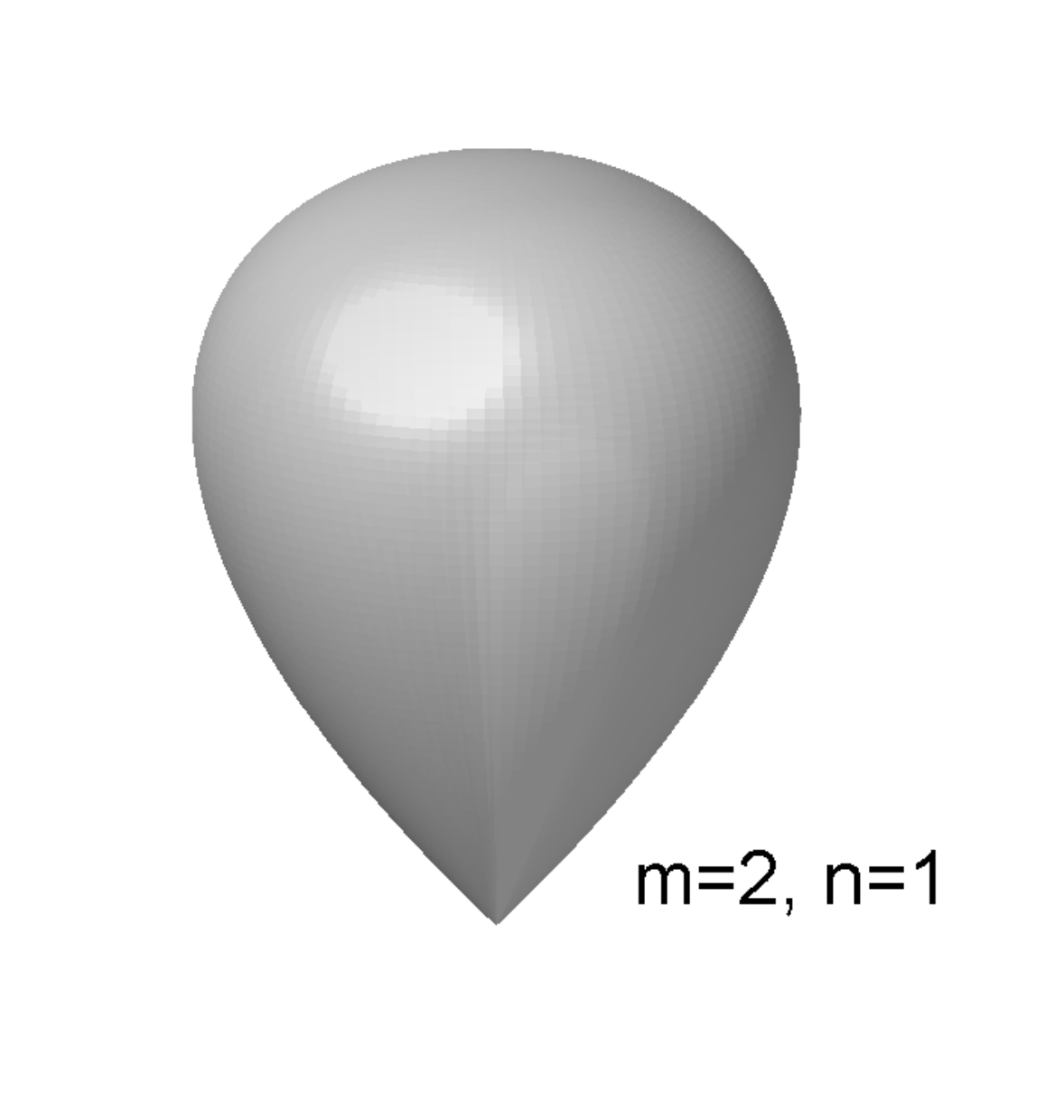}
\hspace*{-4mm}
\includegraphics[width=30mm]{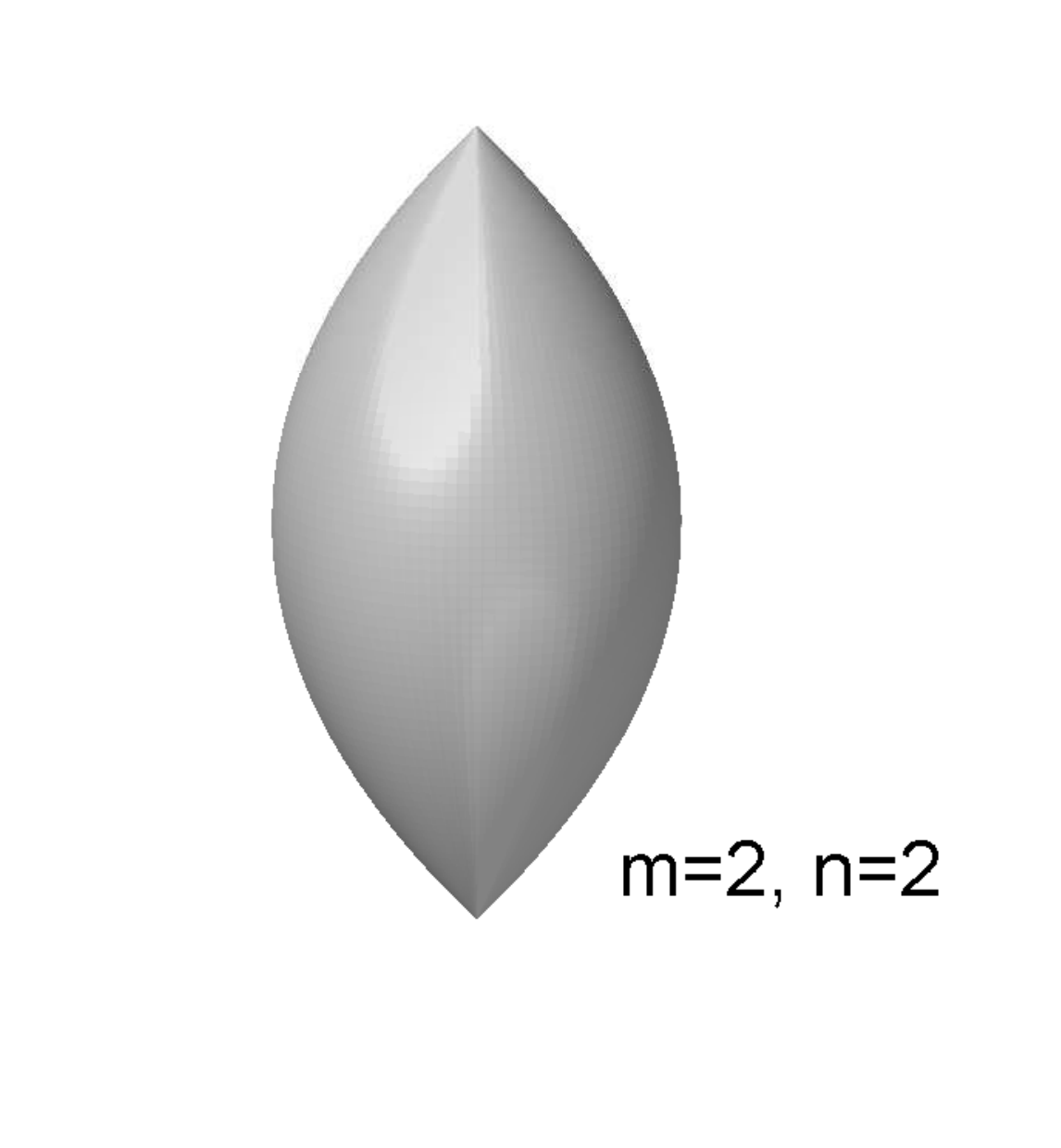}
\hspace*{-8mm}
\includegraphics[width=30mm]{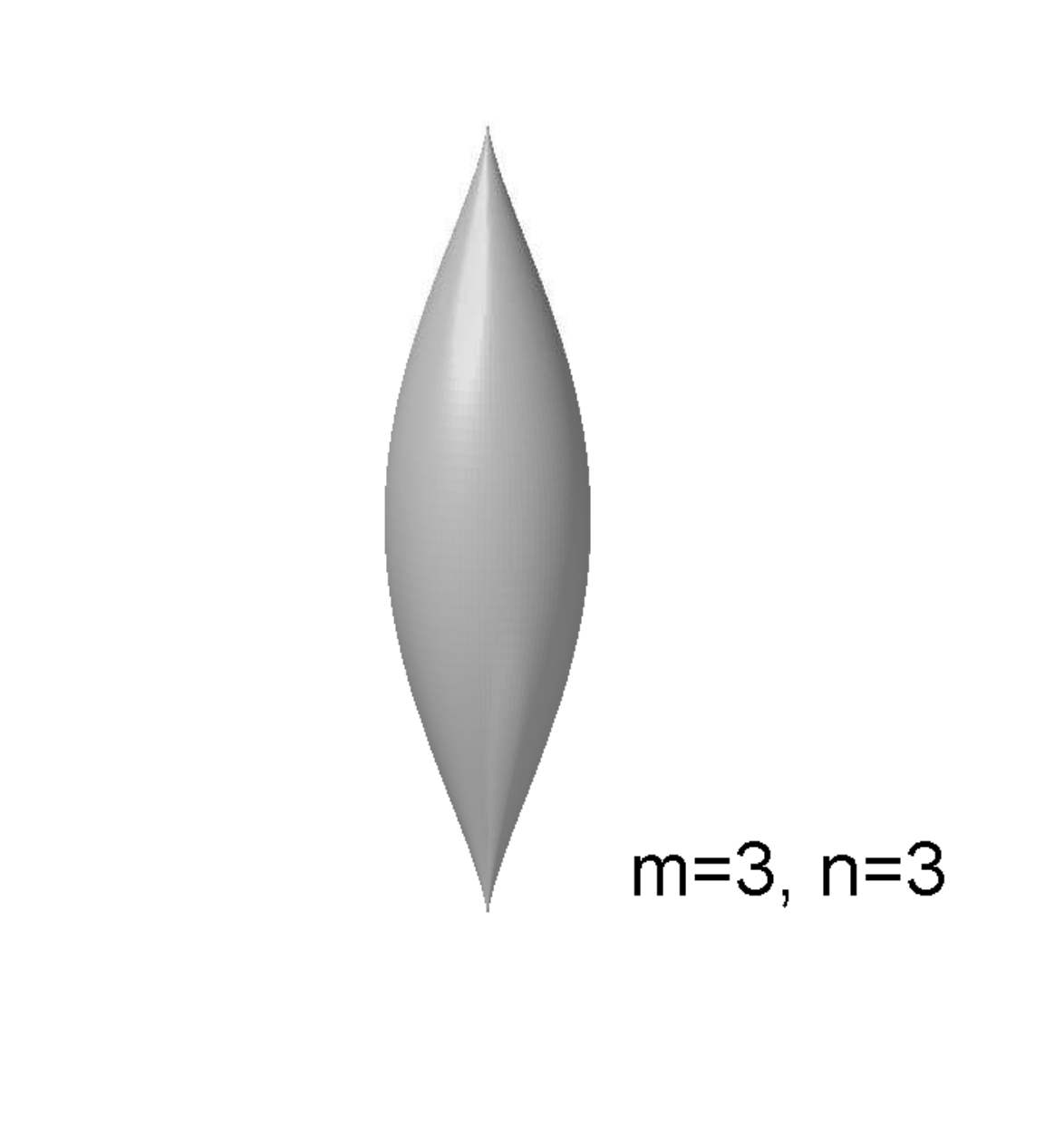}
\includegraphics[width=30mm]{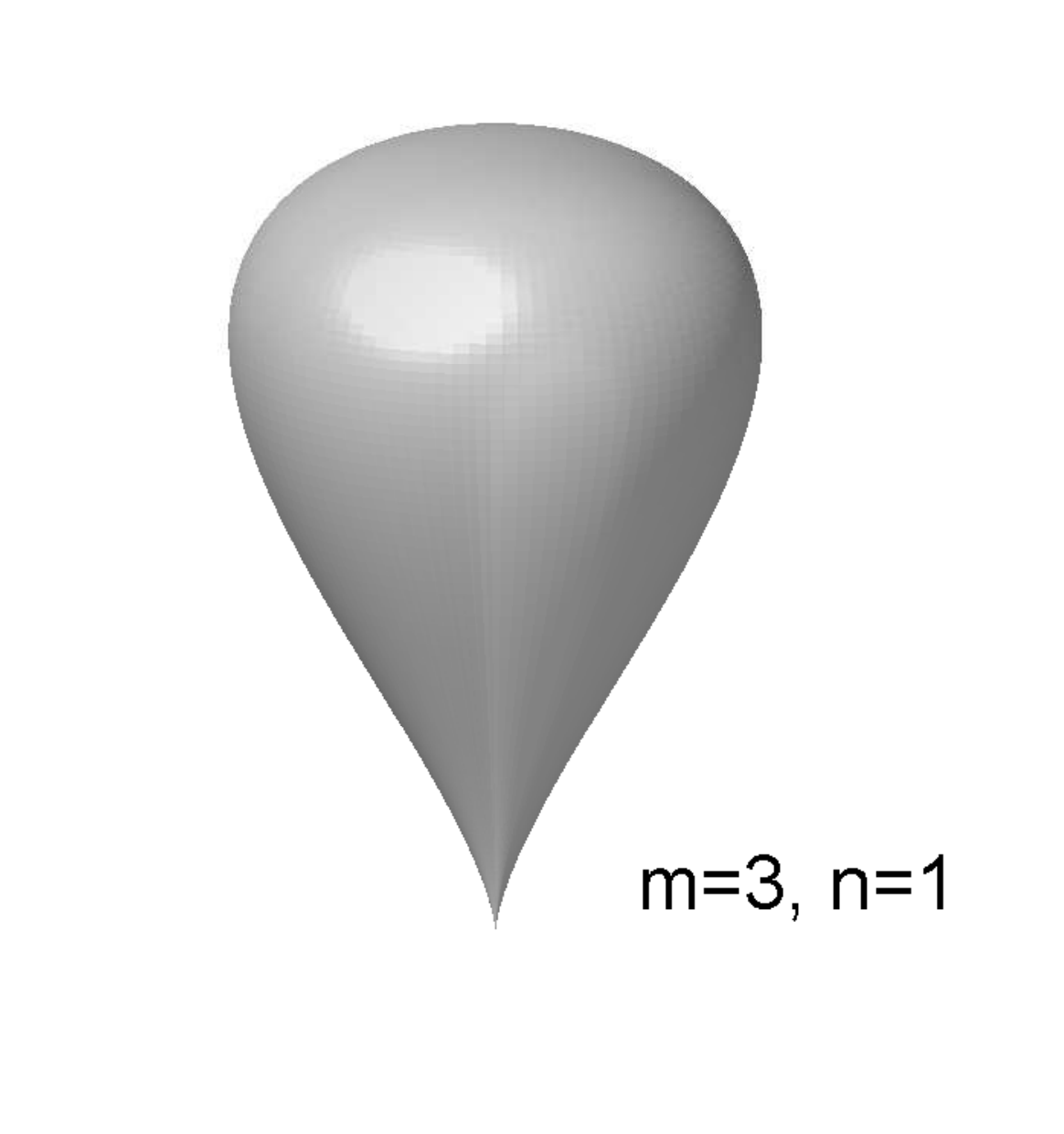}
\includegraphics[width=30mm]{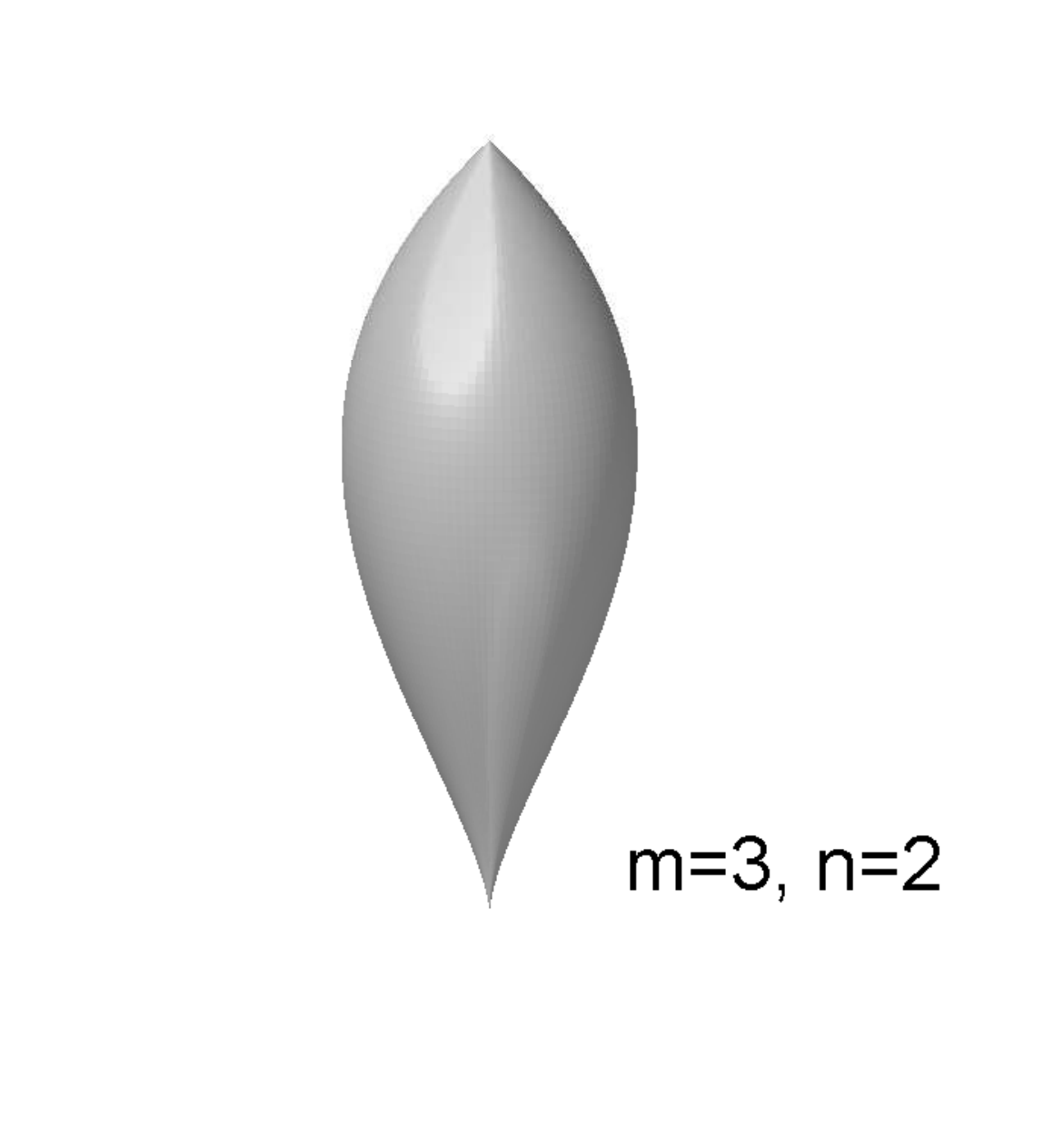}
\hspace*{-4mm}
\end{center}
\caption{\label{fig-shapes} (Color online) Kummer shapes (\ref{kcasi-1}) for selected values of $m$ and $n$.}
\end{figure}

The vector $\bbs=(\kkx,\kky,\kkz)$ evolves in time according to Hamiltonian dynamical equations we shall discuss later, keeping the Casimir function
constant, $C(\bbs)=C$, where the value $C$ can be chosen equal to zero.
An immediate consequence is
the restriction of the dynamics to the orbit manifold
\begin{eqnarray}
\kkx^2+\kky^2&=&-g(\kkz)=r^2(\kkz)\nn\\
&=&
m^{2-n}n^{2-m}\,\big(\tfrac12+\kkz\big)^m\big(\tfrac12-\kkz\big)^{n},
\label{kcasi-1}
\end{eqnarray}
 i.e.~a surface of revolution with a $\kkz$-dependent radius $r(\kkz)$.
 Following Holm, these surfaces will be denoted as Kummer shapes 
 based on previous work by Kummer (see \cite{Holm11a,Kumm81,Kumm86,Kumm90,Holm12}),
which generalizes the Bloch sphere 
\begin{eqnarray}
\label{kcasi11}
\kkx^2+\kky^2=r^2(\kkz)=\tfrac{1}{4}-\kkz^2
\end{eqnarray}
for $(m,n)=(1,1)$ to polynomial algebras. Figure \ref{fig-shapes} shows some examples of these shapes for different values of $n$ and $m$.

\begin{figure}
\begin{center}
\includegraphics[width=40mm]{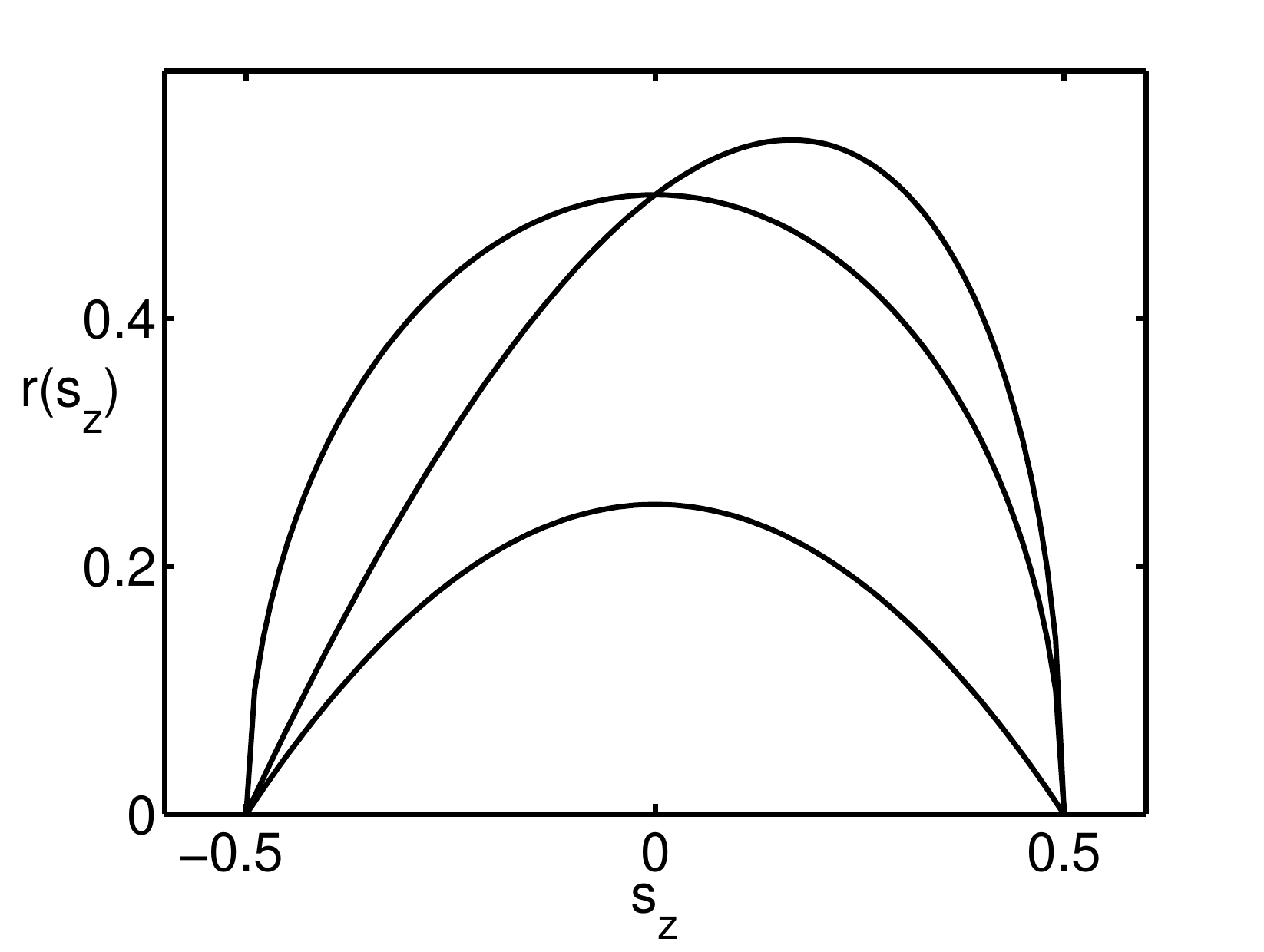}
\includegraphics[width=40mm]{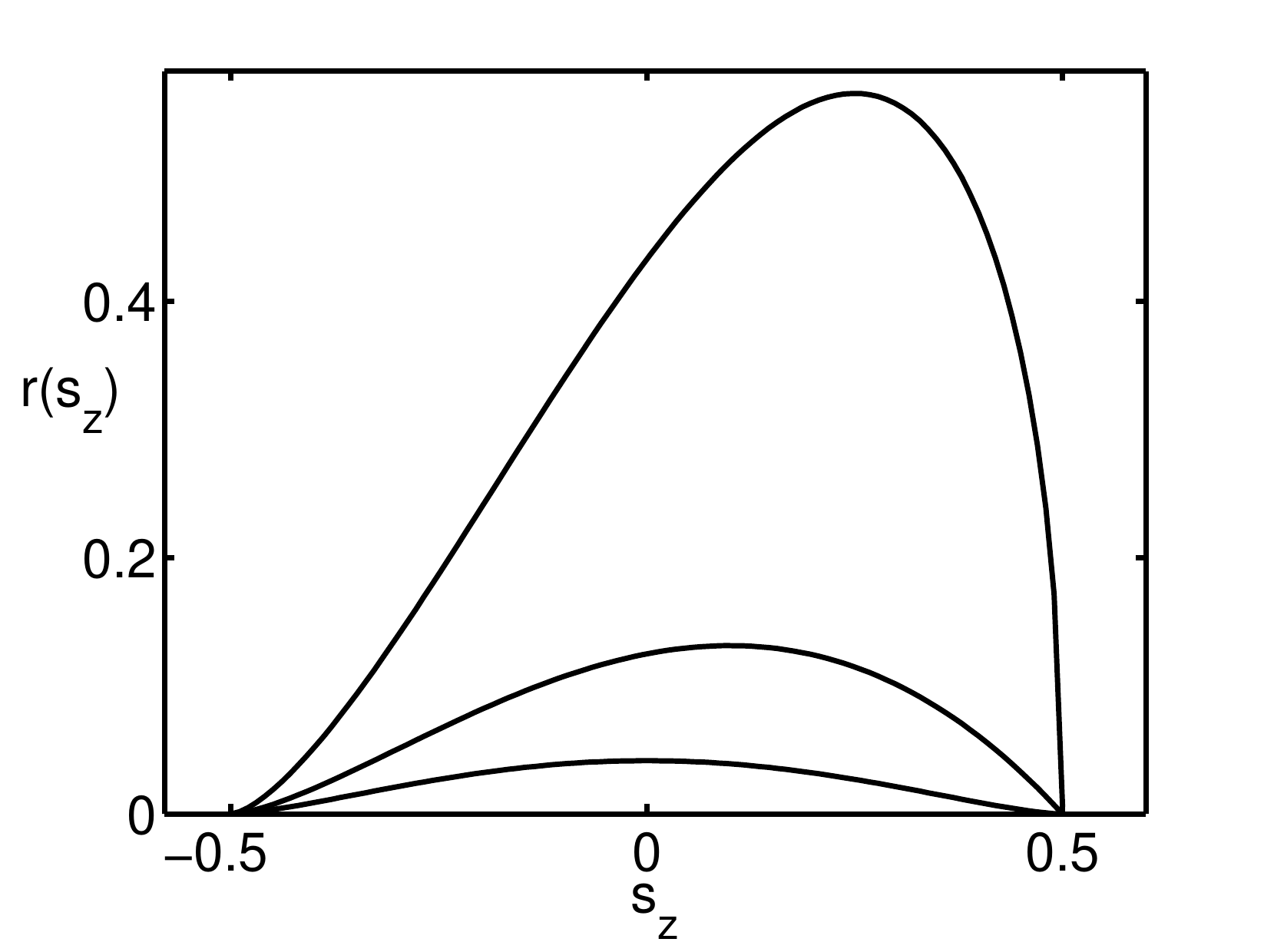}
\end{center}
\caption{\label{fig-radius} Radius $r(\kkz)$ of the Kummer surface (\ref{kcasi-1}) for 
$(m,n)=(1,1)$, $(2,1)$, $(2,2)$ (left) and $(3,1)$, $(3,2)$, $(3,3)$ (right).}
\end{figure}

These Kummer shapes are manifolds with the possible exceptions of the
poles  $\bbs_\pm=(0,0,\pm\tfrac12)$. Here the surface is smooth 
at the north pole $\bbs_+$ for $n=1$, and at the south pole $\bbs_-$ for $m=1$.
For $n\ge2$ or $m\ge 2$ the surfaces are pinched at these points, where we have a 
tip for $m$ or $n$ equal to $2$ and a cusp for larger values. Figure \ref{fig-radius} shows
the radius $r(\kkz)$ as a function of $\kkz$ for selected values of $m$ and $n$. The slope of the
radius $r(\kkz)$ at the north pole $\bbs_+$ is infinite for $n=1$, and the same holds for the south pole for $m=1$. For $m=2$ the
slope at $\bbs_-$ is equal to $2^{1-n/2}$, i.e.~$\sqrt{2}$ for $n=1$, $1$ for $n=2$ and $1/2$
for $n=4$. For $n=2$ the slope at  $\bbs_+$ is equal to $2^{1-m/2}$. For $m,n> 2$ the slope at the
poles is zero.

Let us recall that the dynamics generated by the Hamiltonian
\begin{eqnarray}
\label{ham-cl}
H=v\kkx+\epsilon\kkz
\end{eqnarray}
follows the equations of motion $\dot s_j=\big\{s_j,H\big\}$, i.e.
\begin{eqnarray}
\dot \kkx&=&-\epsilon \kky\, ,\nn\\
\dot \kky&=&\epsilon \kkx-v\kfmn\, , \label{eqn-cl} \\
\dot \kkz&=&v \kky\nn,
\end{eqnarray}
and the conservation
of $C(\kkx,\kky,\kkz)$ restricts the orbit to the Kummer surface (\ref{kcasi-1}),
and in addition by the
conservation of energy, so that the orbits are geometrically given by the intersection
of  the Kummer shape (\ref{kcasi-1}) with the surface $H(\kkx,\kky,\kkz)=E$, 
which is a plane for the Hamiltonian (\ref{hamil-cl}).

As already pointed out in Holm's book \cite{Holm11a} as well as in
\cite{Odzi14}, the Kummer dynamics
can be formulated in terms of the Nambu-Poisson bracket, 
a Lie bracket which also satisfies the Leibnitz relation
\cite{Holm11b}. Using the relation (\ref{g-f-cl}) between $f(\kkz)$ and
 $g(\kkz)$, one can rewrite the equations of motions in terms of the Nambu bracket
\begin{eqnarray}
\label{b-nambu}
\big\{A,B\big\}_{C}=\tfrac12 \nabla C\cdot\big(\nabla A\times \nabla B\big),
\end{eqnarray}
where  $\tfrac12 \nabla C=(\kkx,\kky,f(\kkz))$ is the gradient of the
Casimir function given in (\ref{kcasi-2}), in the convenient form
\begin{eqnarray}
\label{eq-nambu}
\dot A=\big\{A,H\big\}_{C},
\end{eqnarray}
which  immediately reveals the conservation of both the Hamiltonian $H$
and the Casimir function $C$. In addition, the equation of motion for the vector
$\bbs$ can be written as 
\begin{eqnarray}
\label{eq-nambu-s}
\dot{\bbs}=\big\{ \bbs ,H\big\}_{C}=\tfrac12 \nabla C\times \nabla H\,.
\end{eqnarray}
Alternatively, one can describe the dynamics in terms of canonical variables
$p$ and $q$, where $p\in[-\frac{1}{2},\frac{1}{2}]$ is equal to $\kkz$ and $q\in[0,2\pi]$ is the angle in the $\kkx$,
$\kky$ plane \cite{Grae15}:
\begin{eqnarray}
\label{eq-pq}
\kkz=p\ ,\quad \kkx=r(p)\,\cos q\ ,\quad
\kky=r(p)\,\sin q 
\end{eqnarray}
with radius  (compare (\ref{kcasi-1}))
\begin{eqnarray}
\label{eq-r(p)}
r(p)\!=\!r_0\big(\tfrac12\!+\!p\big)^{m/2}\big(\tfrac12\!-\!p\big)^{n/2},
\  r_0\!=\!\frac{1}{\sqrt{m^{n-2}n^{m-2}}}.
\end{eqnarray}
Then the dynamics is given by the standard Poisson bracket
$\{A,B\}=\partial_p A\partial_q B-\partial_q A\partial_p B$ and the
Hamiltonian
\begin{eqnarray}
\label{eq-Hpq}
H(p,q)=\epsilon p+vr(p)\,\cos q\,.
\end{eqnarray}
Note that in this formulation the restriction to the Kummer surface
(\ref{kcasi-1}) is immediately obvious. 

Furthermore this canonical description can be conveniently used as a basis for a semiclassical
WKB-type quantization recovering the individual multi-particle energy eigenvalues and eigenstates that we shall perform in Section \ref{sec_semi}.

%
\subsection{Fixed points}
\label{s-fixed}
Important for the organization of the dynamics both in classical and quantum
mechanics are the fixed points of the motion. The fixed points are found at
$\kky=0$ and $\epsilon \kkx=v\kfmn$. Squaring this equation, using the constraint (\ref{kcasi-1}) and inserting the expression (\ref{kFasy}) for $\kfmn$ this can be rewritten as a condition for the $s_z$ coordinate of the fixed point
\begin{eqnarray}
&&4 \frac{\epsilon^2}{v^2} m^{n-2}n^{m-2}\big(\tfrac12\!+\!\kkz\big)^{m}\big(\tfrac12\!-\!\kkz\big)^{n}\\
&&=\left[n\big(\tfrac12\!+\!\kkz\big)^{m}\big(\tfrac12\!-\!\kkz\big)^{n-1}-m\big(\tfrac12\!+\!\kkz\big)^{m-1}\big(\tfrac12\!-\!\kkz\big)^{n}\right]^2.\nn
\end{eqnarray}
Thus, for $n>1$ the north pole is always a fixed point, independently of the parameter values, and for $m>1$ the same holds for the south pole. 

To find the remaining fixed points, we rearrange to find the condition
\begin{eqnarray}
\label{eq-fixed}
&&16 \epsilon^2 m^{n-2}n^{m-2}\\
&&=v^2\big(\tfrac12\!+\!\kkz\big)^{m-2}\big(\tfrac12\!-\!\kkz\big)^{n-2}\,
\big(2(n+m)\kkz+n-m\big)^2.\nn
\end{eqnarray}
In the special case $m=n$ this simplifies to
\begin{eqnarray}
\label{eq-fixed-sym}
\epsilon^2 m^{2m-6}
=v^2\big(\tfrac{1}{4}-\kkz^2\big)^{m-2}\kkz^2\,.
\end{eqnarray}
The real roots of the polynomial (\ref{eq-fixed}) with $-\tfrac12\le \kkz\le +\tfrac12$, $\kkx=\frac{v}{\epsilon}\kfmn$, and $\kky=0$, yield the fixed points, in addition to those at the poles for $m,n>1$. That is, the total number of fixed points for a given $n$ and $m$ is bounded by $n+m$. As we shall see in what follows, however, for $n+m\ge 6$ the maximal number of fixed points is six.  

At a fixed point (apart from those at the poles for $n$ or $m$ larger than one) the energy plane $E=v\kkx+\epsilon \kkz$ is tangential to the Kummer surface, i.e.~the slope of the straight line $\kkx=(E-\epsilon\kkz)/v$ must be equal to the slope
of $r(\kkz)$. Thus, we need to determine at how many points the slope of $r(\kkz)$ can have a prescribed value (it is instructive to have a look at figure \ref{fig-radius}). For this purpose it is useful to divide the Kummer surface in a southern and northern part along the line of maximal radius (the `equator'). The value of $m$ determines the qualitative behaviour of the slope on the southern part, the value of $n$ that on the northern part. Let us consider the southern part in dependence on $m$. For $m=1$ the slope at the south pole is infinite and decreases monotonically to zero at the equator. That is, there is exactly one value of $s_z$ corresponding to each prescribed value of the slope, and thus we have one fixed point on the southern part of the Kummer shape. For $m=2$ the south pole is a fixed point for all parameter values. The slope at the south pole is equal to $2^{1-n/2}$ and decreases monotonically to zero at the equator. Thus, there is exactly one point at which the energy plane is tangential to the southern part of the Kummer shape for $\epsilon^2/v^2\le  2^{2-n}$ and none otherwise. In total we thus have either one or two fixed points on the southern part. For $m\ge 3$ we always have the same scenario: The south pole is a fixed point for all parameter values, the slope at the south pole is zero, with increasing $s_z$ it increases to a maximum at the point of inflection, after which it decreases to zero at the equator. Thus, for values of $|\epsilon/v|$ smaller than the maximal value of the slope there are two points on the southern part of the Kummer shape at which the energy plane is tangential to the shape, for larger values there is none. In total there are thus either one or three fixed points on the southern part of the Kummer shape. The same holds for the northern part in dependence on $n$. In summary, the maximal number of fixed points for given values of $n$ and $m$ is given by $\min(n+m,6)$. 

The character of the fixed points can be determined from the eigenvalues of the Jacobi matrix
\begin{eqnarray}
\label{eig-jac}
\lambda_\pm=\pm\sqrt{-\epsilon^2-v^2f'(\kkz)}\,,
\end{eqnarray}
which are a complex conjugate pair for a center, in the vicinity of which the motion is a rotation with
frequency 
\begin{eqnarray}
\label{fix-om}
\omega=\sqrt{\epsilon^2+v^2 f'(\kkz)},
\end{eqnarray}
or a pair of
real numbers with different signs for a saddle point. 

At critical parameter values the number of fixed points changes. This
can happen in two ways at $\epsilon_c$: \\[2mm]
(i) Two fixed points can coalesce and disappear. This saddle-node bifurcation 
must necessarily occur
at the inflection point $\kkz$ of $r(\kkz)$ and the critical value of $\epsilon$
is given by the slope at this point: $\epsilon_c=\pm v r'(\kkz)$. \\[2mm]
(ii) Fixed points can enter or leave the system
at the poles in a transcritical bifurcation \cite{Grae15}. Because the slope of $r(\kkz)$
at the poles is zero for $m,n>2$, this can (for $\epsilon\ne 0$) only happen if $m$ or $n$ is equal to $2$,
for $m=2$ at the south pole for $\epsilon_c=v \,2^{1-n/2}$ and
for $n=2$ at the north pole for $\epsilon_c=v\,2^{1-m/2}$. A stability analysis shows that
here a center at the pole changes into a saddle and a new center appears moving 
away form the pole.  \\[2mm] 
One should note that the sum of the Poincar\'e indices (centers have index +1,
saddles index -1,  see, e.g., \cite{Arno06}) remains constant on the Kummer surface
for bifurcations of type (i), whereas it changes for type (ii).
\section{Mean-field and many-particle correspondence}
\label{s-corr}
\begin{figure}
\begin{center}
\includegraphics[width=40mm]{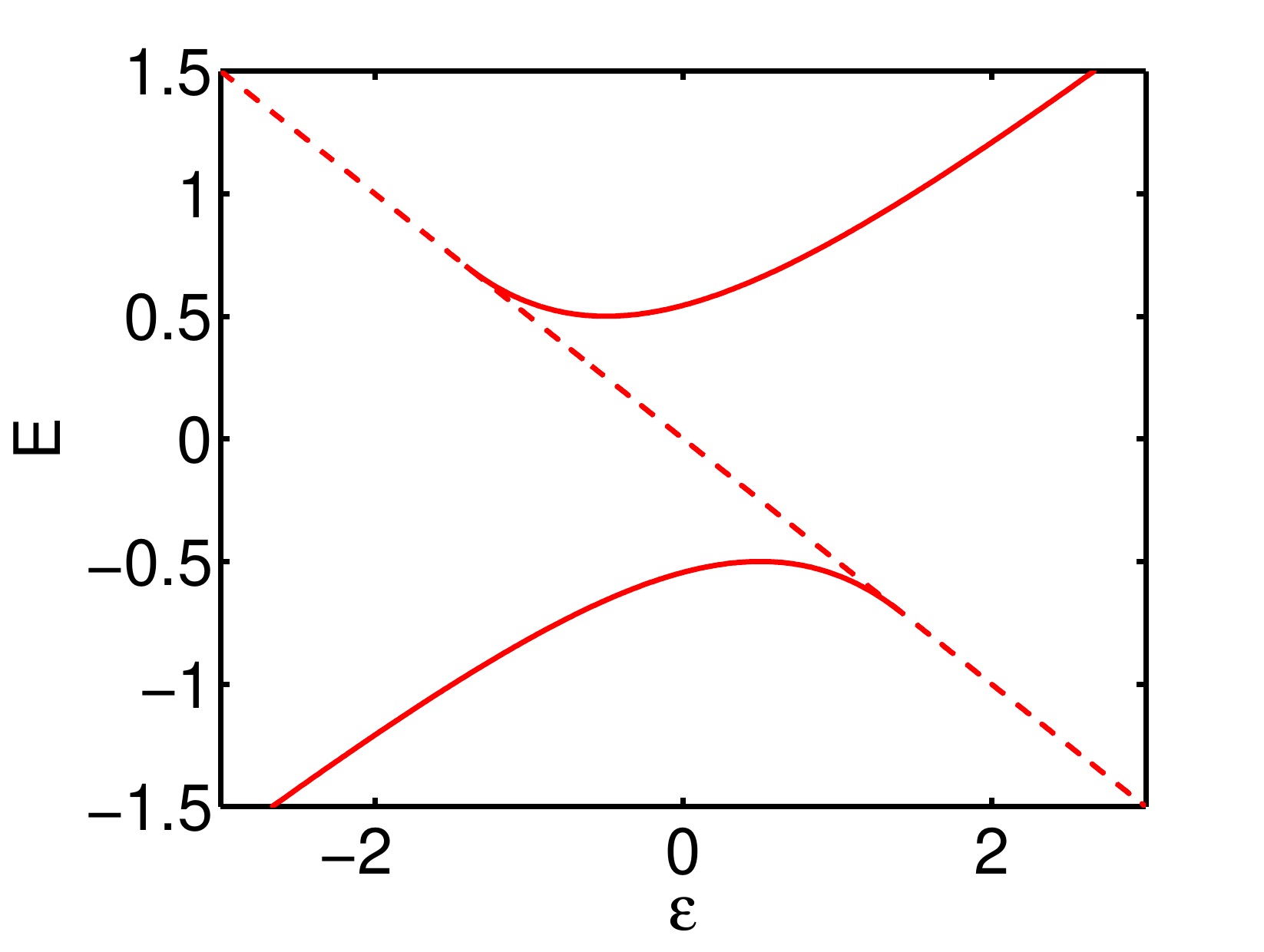}
\includegraphics[width=40mm]{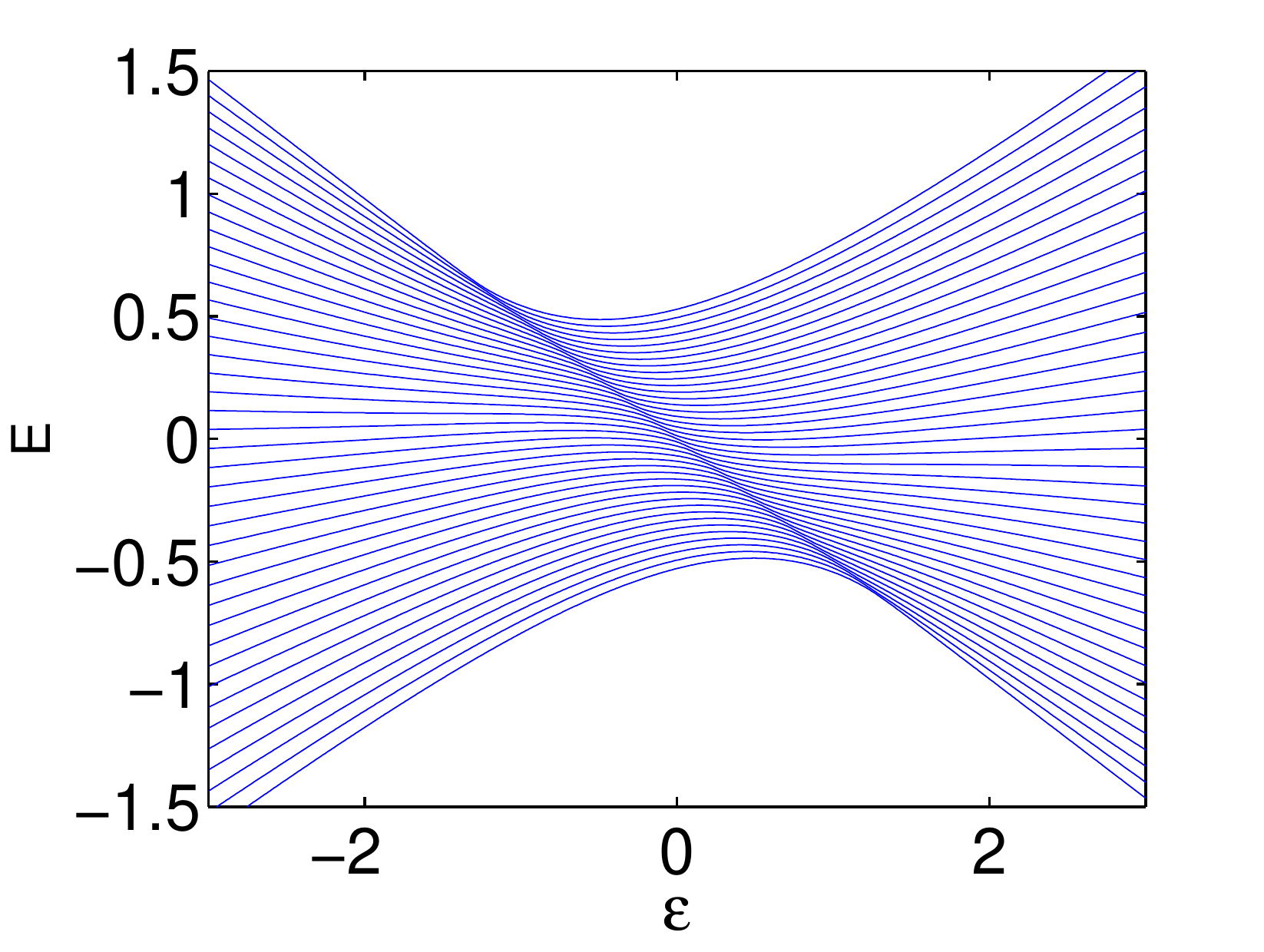}
\end{center}
\caption{\label{fig-E21} (Color online) $(m,n)=(2,1)$\,:\, Mean-field fixed point energies (red) and
(scaled) many-particle energies (blue) in dependence of $\epsilon$ for $v=1$ and $N=80$ particles.}
\end{figure}

\begin{figure}
\begin{center}
\includegraphics[width=40mm]{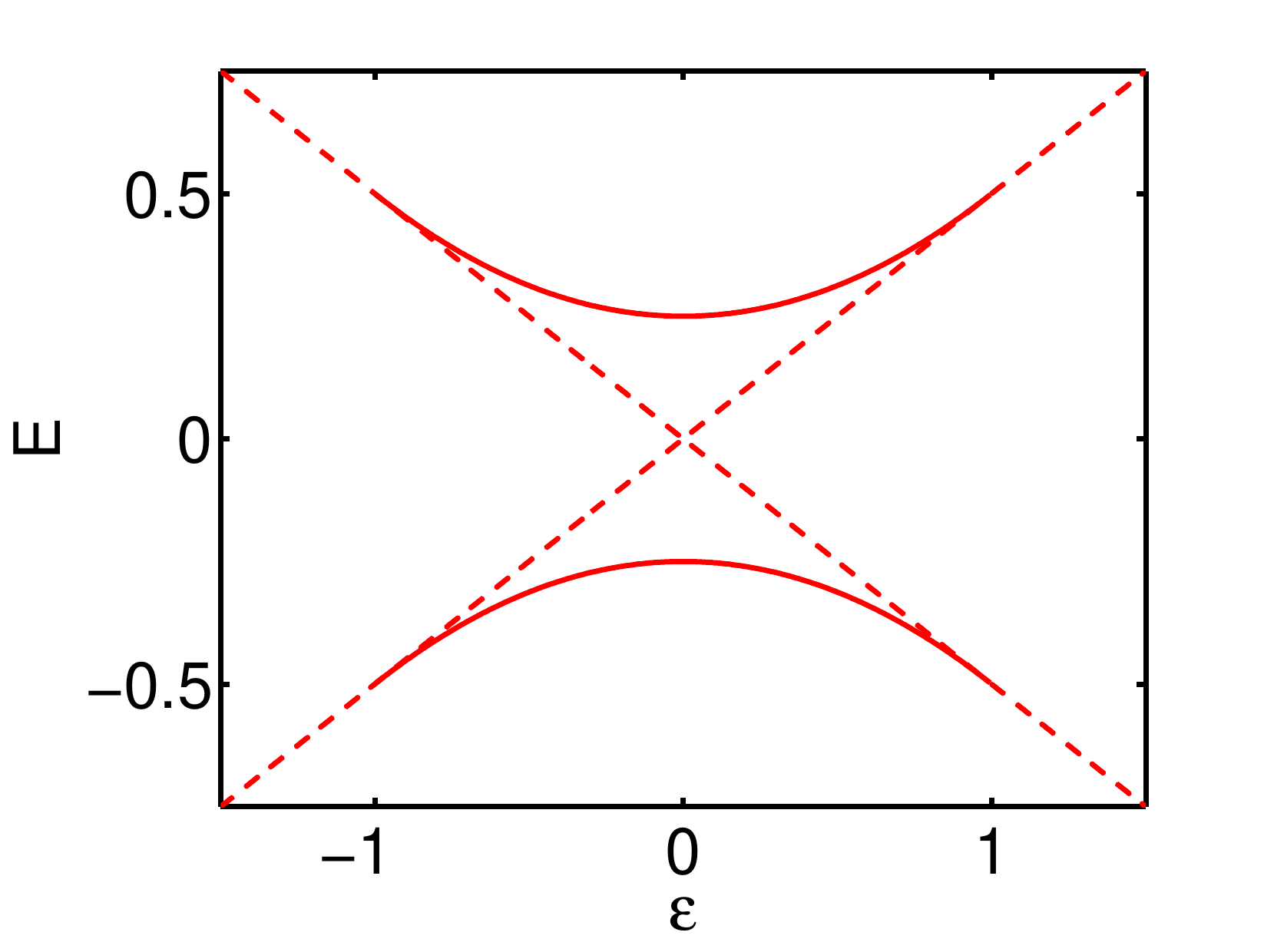}
\includegraphics[width=40mm]{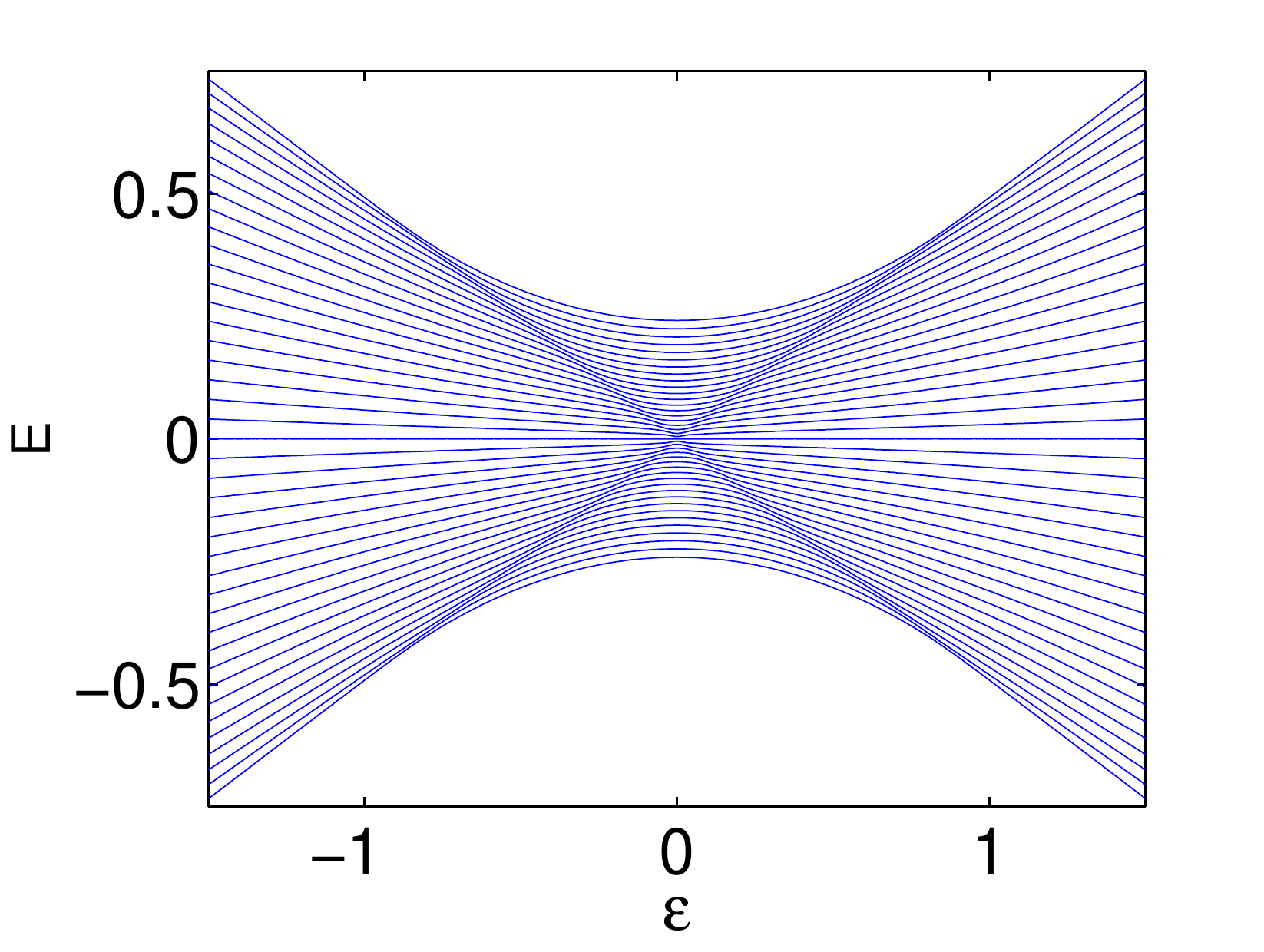}
\end{center}
\caption{\label{fig-E22} (Color online)  $(m,n)=(2,2)$\,:\, Mean-field fixed point 
energies  (red) and (scaled)  many-particle energies (blue)
in dependence of $\epsilon$ for $v=1$ and $N=160$ particles.}
\end{figure}

\begin{figure}
\begin{center}
\includegraphics[width=40mm]{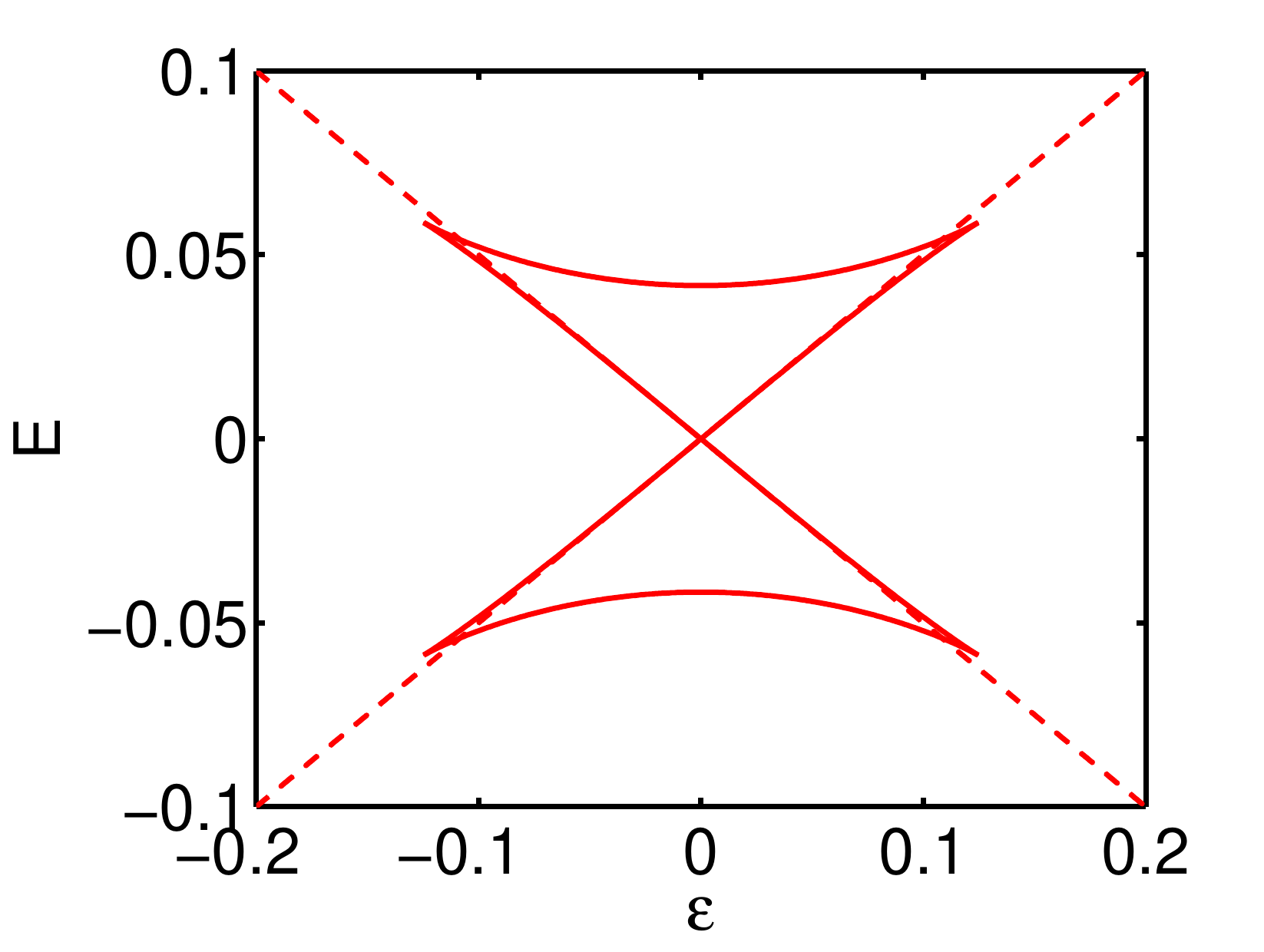}
\includegraphics[width=40mm]{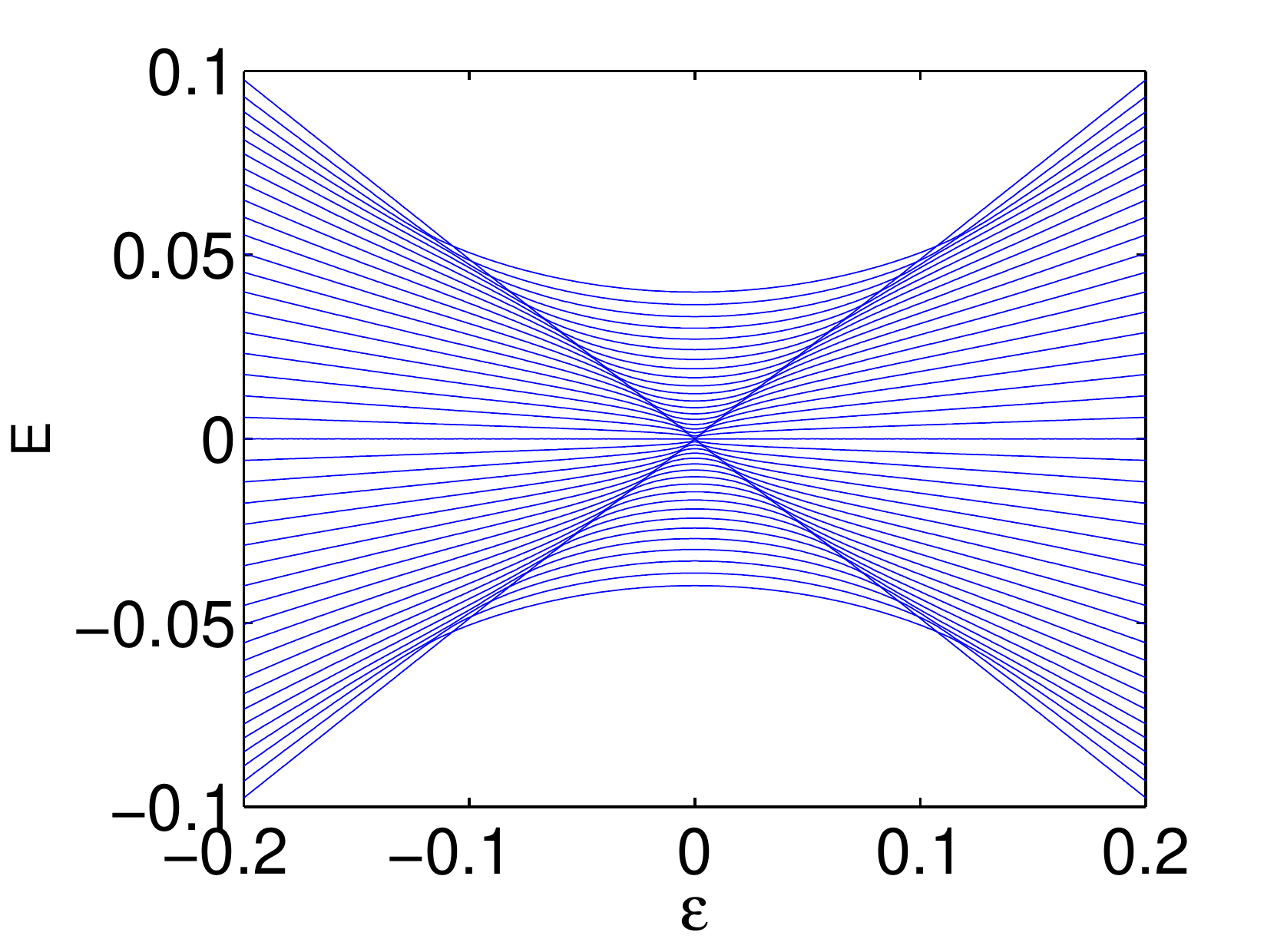}
\end{center}
\caption{\label{fig-E33} (Color online) $(m,n)=(3,3)$\,:\, Mean-field fixed point energies (red) and
(scaled) many-particle energies (blue) 
in dependence of $\epsilon$ for $v=1$ and $N=360$ particles.}
\end{figure}

Let us now discuss the correspondence between quantum many-particle eigenvalues and mean-field dynamics for some cases in more detail. We will use the notation $p,q$ introduced at the end of section
\ref{s-dyn}.
The allowed mean-field energy interval is bounded by the maximum and minimum of the classical Hamiltonian on the phase space, that is the maximum and minimum of the energies
\begin{eqnarray}
\label{E-fixed}
E_f=\frac{v^2}{\epsilon}\,f(p_f) + \epsilon p_f
\end{eqnarray}
at the fixed points $p_f$, that is, the poles or the real solutions of the polynomial (\ref{eq-fixed}) in the interval $[-\tfrac12,+\tfrac12 ]$. For large values of $|\epsilon|$, in the supercritical regime, where we have only two fixed points (at the poles for $n,m>1$), the mean-field energy interval is $E_-<E<E_+$. In the subcritical regime below the critical value(s) of $\epsilon$, there are additional fixed points which correspond to stationary values of the energy, the global extrema are not located at the poles in this case.
The extrema of the mean-field energy are upper and lower bounds for the many-particle energies (rescaled by $\eta$). The additional stationary values do not correspond to individual many-particle eigenvalues, but mark lines along which many-particle eigenvalues accumulate.  To demonstrate this correspondence we show examples of the many-particle spectrum together with the mean-field stationary energies in dependence on the parameter $\epsilon$ for several values of $n$ and $m$, corresponding to the examples depicted in figure \ref{fig-shapes}, in figures \ref{fig-E21}, \ref{fig-E22}, and \ref{fig-E33}. 

It is worthwhile to note that in all cases the many-particle eigenvalues are non-degenerate
so that all apparent crossings are avoided as already observed before 
for an atom-molecule conversion system \cite{Zhou03}. This is simply a consequence of the
fact that the Hamiltonian is tridiagonal (see section \ref{s-fock}) and hence can only
have eigenvalue degeneracies if all off-diagonal elements vanish \cite{Wilk65}. 
\\[1mm]
\noindent
(1) For $(m,n)=(1,1)$, i.e.~for the dynamics on the Bloch sphere, we
have $f(p)=p$ and 
the two fixed points are at $p_\pm=\pm 1/(2\sqrt{1+v^2/\epsilon^2})$, $q_+=0$,
$q_-=\pi$, 
which are both centers with frequency $\omega=\sqrt{\epsilon^2+v^2}$ and an
energy 
\begin{eqnarray}
E_\pm=H(p_\pm,q_\pm)=\pm \tfrac{1}{2}{\sqrt{\epsilon^2+v^2}}\,,
\end{eqnarray}
which are upper and lower bounds  for the many-particle eigenvalues (rescaled by $\eta$). 
\\[2mm]
\noindent
(2) The case $(m,n)=(2,1)$, describing the dissociation and association of diatomic molecules, has been
analyzed in \cite{Grae15,Pere11}. Here the Kummer surface 
\begin{eqnarray}
&&\kkx^2+\kky^2=r^2(p)\nn\\
&&=2\big(\tfrac{1}{2}+p\big)^2\big(\tfrac12-p\big)
=\tfrac{1}{4}+\tfrac{1}{2}p-p^2-2p^3\label{kummer21}
\end{eqnarray}
has the shape of a teardrop (see figure \ref{fig-shapes}). As discussed before, at the north
pole it is smooth and there is a tip at the south pole. The slope of $r(p)$ at the south pole
is equal to $\sqrt 2$, which is the critical value of $\pm \epsilon/v$. Therefore
the supercritical region with only two fixed points is given by $|\epsilon/v| >\sqrt{2}$ (see also \cite{Grae15,Pere11}).
With $f(p)=-\tfrac{1}{4}+p+3p^2$ the equations of motion
written in terms of the $s_j$ are
\begin{eqnarray}
\label{eqn-cl-21}
\dot s_x&=&-\epsilon \kky\,,\nn\\
\dot s_y&=&\epsilon \kkx-v\big(\!-\!\tfrac{1}{4}\!+\!\kkz\!+\!3\kkz^2\big),\\
\dot s_z&=&v \kky\nn
\end{eqnarray}
and the equation for the value of $p$ at the fixed points reads
\begin{eqnarray}
\label{fp-21}
\big(\tfrac12+p\big)\big(9v^2p^2+(2\epsilon^2-3v^2)p+\tfrac{v^2}{4}-\epsilon^2\big)=0
\end{eqnarray}
with the expected solution $p=-\tfrac12$ and two solutions of the remaining quadratic equation, one
of which is in the interval $-\tfrac12\le p\le \tfrac12$ for all parameter values, while the second one only lies in this physical region in the subcritical case  \cite{Grae15,Pere11}.
The $x$-component can be determined by $\kkx=vf(p)/\epsilon$.
At the south pole the slope of $f(p)$ is equal to $-2$, so that the eigenvalues (\ref{eig-jac}) of the stability matrix
are $\lambda=\pm \sqrt{-\epsilon^2+2v^2}$, that is, the fixed point at the south pole is a center in the supercritical case and a saddle point
in the subcritial regime. The remaining one or two fixed points are always centers.  Detailed numerical examples can be found in \cite{Grae15}.
The mean-field energies at the fixed points in dependence on the parameter $\epsilon$ are shown in figure \ref{fig-E21}  
and compared with the quantum eigenvalues for $N=80$ particles (corresponding to a matrix size of $41$), which are clearly organized by the classical 
fixed point energies. (Note that the many-particle energies $E$ must be rescaled by a factor $\eta$ for comparison.)
\\[2mm]
\noindent
(3) For the case $m=n=2$, which can be interpreted as pair-tunneling and which corresponds to the second shape in figure \ref{fig-shapes}, we have
\begin{eqnarray}
\label{fr22}
r(p)=\tfrac{1}{4}-p^2\ ,\quad
f(p)=2p \big(\tfrac{1}{4}-p^2\big).
\end{eqnarray}     
We have two fixed points at the poles, and for $\epsilon^2<v^2$ two additional fixed points with
\begin{eqnarray}
\label{fp-22}
s_z=\pm\frac{\epsilon}{2v}\ ,\quad  \kkx=\frac{v}{\epsilon}\,f(p)=
\pm \frac{1}{4}\Big(1-\frac{\epsilon^2}{v^2}\Big),
\end{eqnarray} 
which are centers. In this parameter region the fixed points at the poles
are saddles. For $\epsilon^2>v^2$ we find only two centers at the poles.
The energy at the fixed points (\ref{fp-22}) is given by
\begin{eqnarray}
\label{E-22}
E_{1,2}=\pm\frac{v}{4}\Big(1+\frac{\epsilon^2}{v^2}\Big).
\end{eqnarray} 
In dependence on $\epsilon$ this is a curve that joins smoothly with the energies $E_{\pm}=\pm \epsilon/2$ of the fixed points at the poles at the critical values $\epsilon=\pm v$. Figure \ref{fig-E22} shows
the mean-field energies at the fixed points and
the quantum eigenvalues for $N=160$ particles (corresponding to a matrix size of $41$ as in the previous example), which are again supported by the classical skeleton of
fixed point energies.\\[2mm]
\noindent
(4) The case $(m,n)=(3,3)$, corresponding to the third shape in figure \ref{fig-shapes}, is more involved. First we have
\begin{eqnarray}
\label{fr33}
r(p)=\tfrac{1}{3}\big(\tfrac{1}{4}-p^2\big)^{3/2}\ ,\quad
f(p)=\tfrac{1}{3}p\big(\tfrac{1}{4}-p^2\big)^2
\end{eqnarray}     
and the fixed points are found from (\ref{eq-fixed-sym}), a 
second order polynomial in $p^2$, as
\begin{eqnarray}
\label{fp-33}
p=\pm\sqrt{\tfrac{1}{8}\big(1\pm \sqrt{1-(8\epsilon/v)^2}\big)}
\ ,\quad  \kkx=\tfrac{v}{\epsilon}\,f(p)
\end{eqnarray} 
for $(8\epsilon)^2<v^2$. Note that here we have four fixed points
in addition to the poles, which is the maximum number possible, as discussed
above. 

Figure \ref{fig-E33} shows the energies at the six fixed points
in dependence of $\epsilon$. The four non-trivial ones trace out
a double swallow tail curve with four cusps at the critical values $\epsilon_c=\pm v/8$
with  energy $E=\pm v/12\sqrt{2}=\pm \sqrt{2}\epsilon/3$, which is slightly smaller than the energy
$\pm \epsilon/2$ at the poles. The fixed points close to the line $\pm \epsilon$
are saddle points, those on the curved lines passing through $E=\pm v/24$ for $\epsilon=0$
are centers. At the cusps the character changes, which can also be seen from the
vanishing of the eigenvalues of in Jacobi matrix (\ref{eig-jac}). Again, as demonstrated
in figure \ref{fig-E33} for $N=320$ particles ($N_{\rm dim}=41$), the classical 
fixed point energies provide a skeleton for
the quantum eigenvalues.\\[2mm]

\begin{figure}
\begin{center}
\includegraphics[width=40mm]{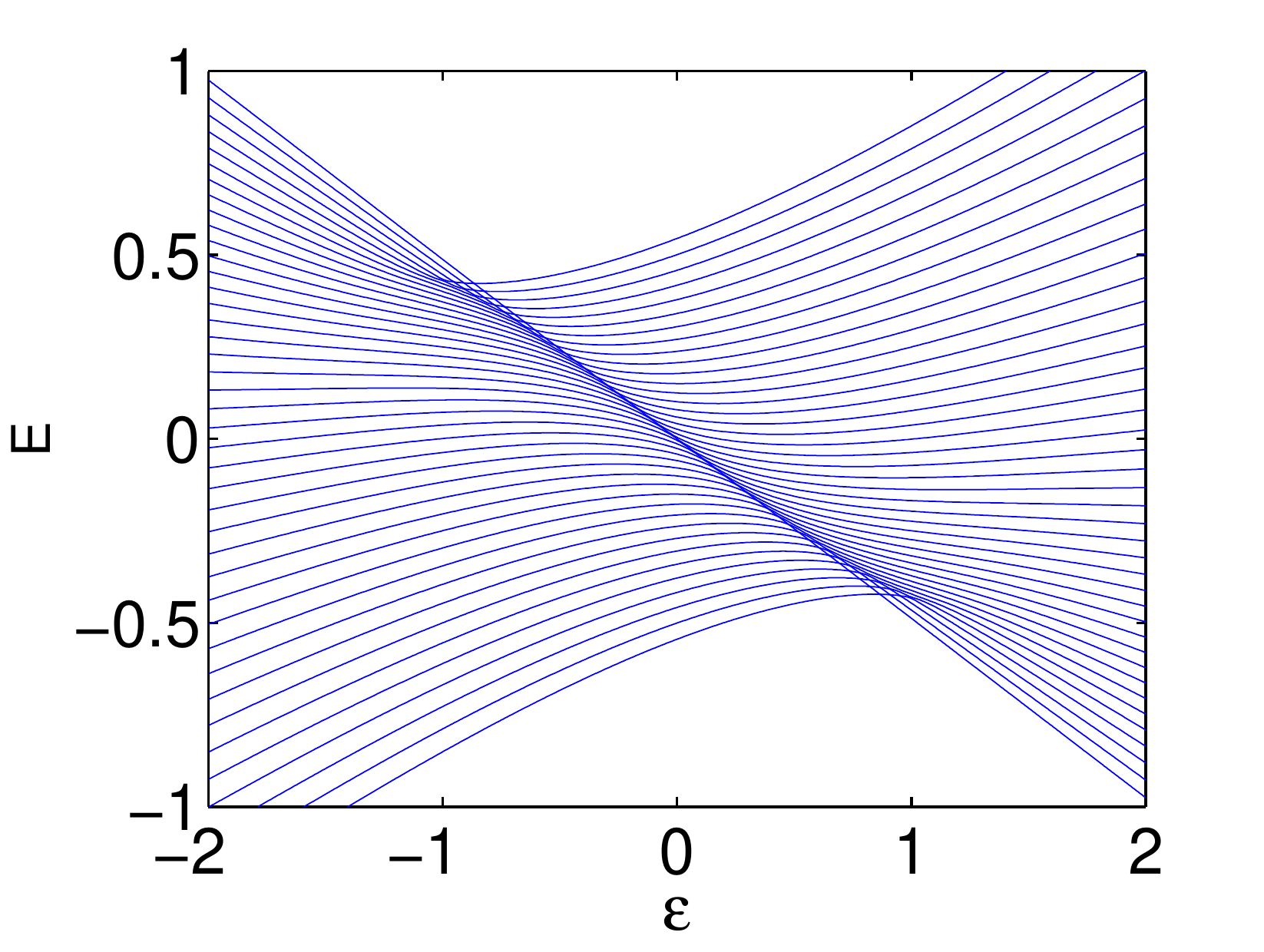}
\includegraphics[width=40mm]{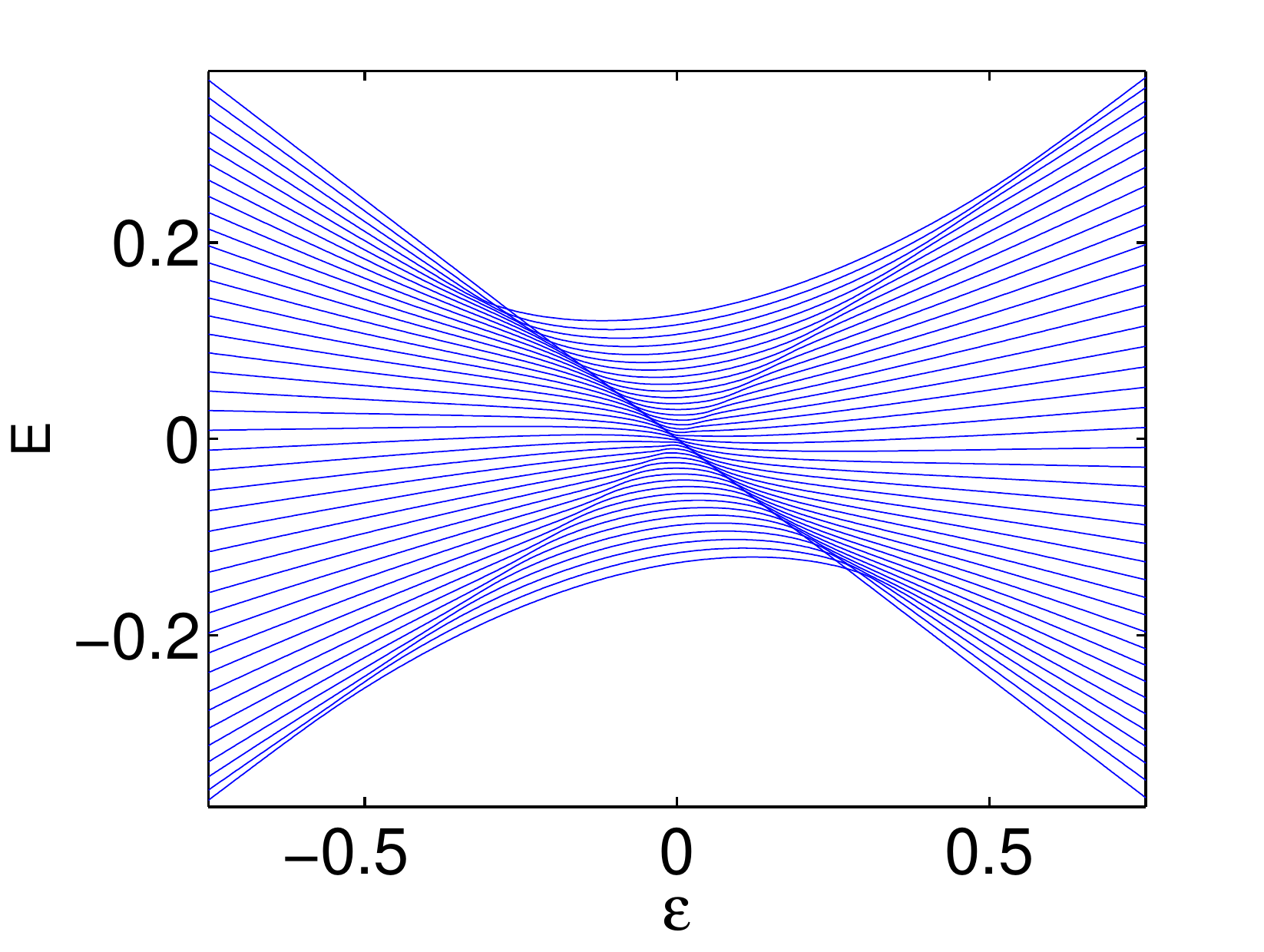}
\end{center}
\caption{\label{fig-E31-32} (Color online) Many-particle  (scaled) energies
in dependence of $\epsilon$ for $v=1$ and $(m,n)=(3,1)$ (left, $N=120$ particles)
and $(m,n)=(3,2)$ (right, $N=240$ particles).}
\end{figure}

\noindent
(5) Finally we will briefly consider the cases $(m,n)=(3,1)$ and $(3,2)$
whose energy eigenvalues are shown in figure \ref{fig-E31-32} for
$N=120$ or $240$ particles. Their structure should be understandable
now without presenting their classical skeleton. 

The figure on the left, for $(3,1)$, is a combination of the structures already 
shown in figures \ref{fig-E21}
and  \ref{fig-E33} for $(m,n)=(2,1)$ and $(3,3)$, respectively. At the
north pole the Kummer surface is smooth and generates no bifurcation. At
the south pole we find a cusp, leading to a cusp singularity as in the case $(3,3)$
showing up in the upper left and lower right of the $(E,\epsilon)$-plane.
 
The figure on the right, for $(3,2)$, also combines features discussed before.
Again we observe the cusps on the upper left and lower right, but here we
also have a tip of the Kummer surface at the north pole, giving rise to a bifurcation
and the additional line $E=\epsilon/2$ as already seen in figures  \ref{fig-E22} 
and \ref{fig-E33}.
\section{Semiclassical quantization and density of states}
\label{sec_semi}
We can recover the many-particle spectrum from the mean-field system from a WKB type quantization condition 
as carried out for $(m,n)=(1,1)$ in  \cite{07semiMP,Niss10,Simo12} and for $(m,n)=(2,1)$ in  \cite{Grae15}. In the case where there is a single classically allowed region for any given energy the quantization condition is given by
\begin{equation}
\label{eq:quantisationCondition}
S\left( \eta E_\nu \right) = 2 \pi \eta \left( \nu+\frac{1}{2} \right),
\end{equation}
where $\nu \in \left\{ 0, 1, 2,\dots, \frac{N}{mn}\right\}$, and where $S(\eta E)$ denotes the phase space area enclosed by the orbit corresponding to the mean-field energy $H=\eta E$.
\begin{figure}
\begin{center}
\includegraphics[width=40mm]{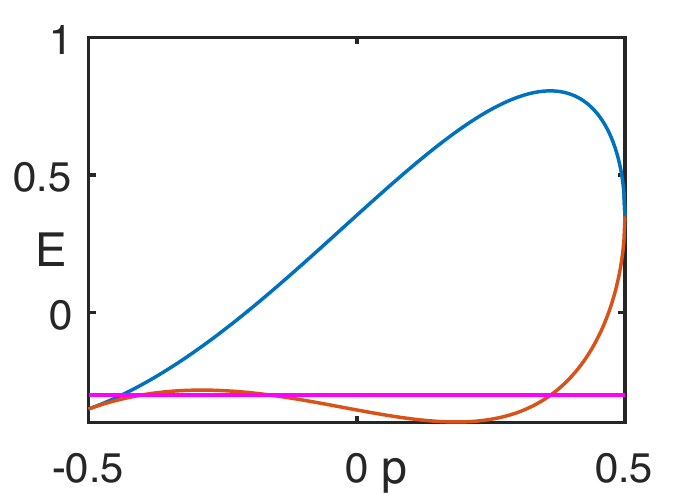}
\includegraphics[width=40mm]{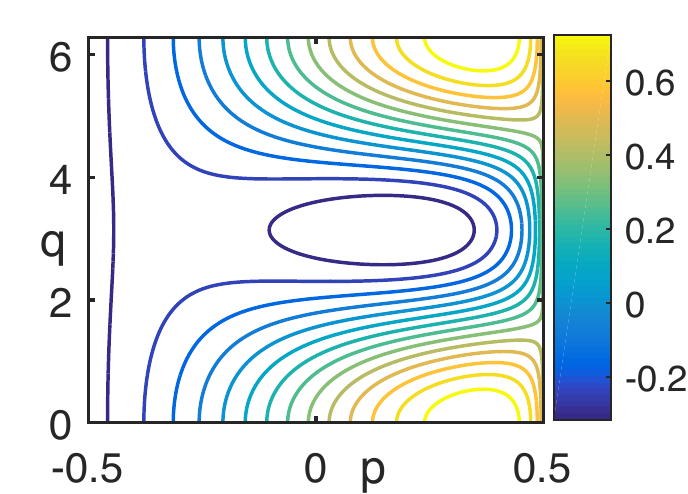}\\
\includegraphics[width=40mm]{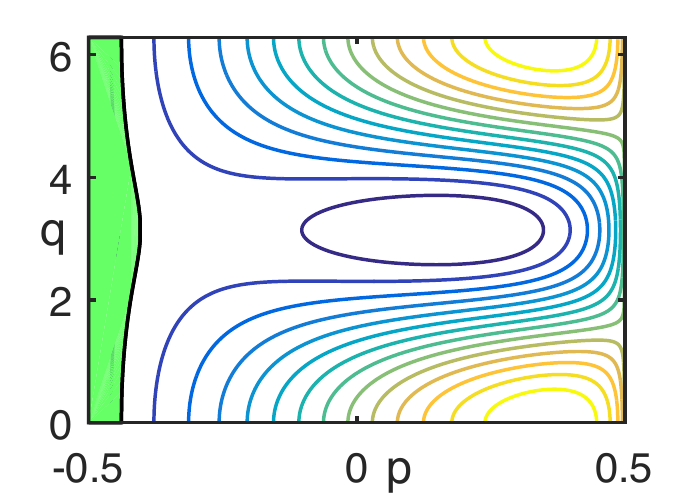}
\includegraphics[width=40mm]{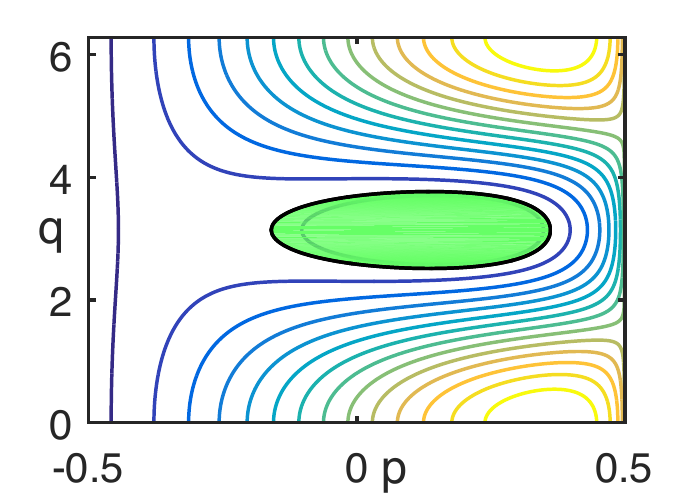}
\end{center}
\caption{\label{fig:quantFig1}(Color online) $(m,n) = (4,1)$ Potential curves $U_\pm$ with energy $E=-\frac{1}{4}$ (top left), contour plot of $H$ in phase space (top right), phase space area in the left region at $E$ (bottom left) and phase space area in the right region at $E$ (bottom right).}
\end{figure}

It is useful to introduce the mean-field momentum `potential functions' $U_\pm(p)$, which are the maximum and minimum curves of the Hamiltonian with respect to the angle variable $q$, given by
\begin{eqnarray}
\label{eq-UpmT}
U_\pm (p)=\epsilon p\pm v\,r(p).
\end{eqnarray}
These potential curves provide lower and upper bounds of the mean-field energy and join at the poles $U_+(\pm\tfrac12)=U_-(\pm\tfrac12)=\pm \tfrac{\epsilon}{2}$. The real valued solutions $p$ of $U_\pm (p) = \eta E$ that fall into the interval $[-\frac{1}{2}, \frac{1}{2}]$ are the turning points of the dynamics. The phase space area can then be calculated from the action integral, 
\begin{equation}
\tilde S(\eta E)=\int_{p_{-}}^{p_+}q\left(p\right)dp,
\end{equation}
between the turning points, with
\begin{equation}
\label{eq:qofp}
q(p)=\arccos \left( \frac{\eta E-\epsilon p}{v r(p)} \right).
\end{equation}
Depending on whether each turning point lies on $U_-$ or $U_+$ the phase space area is given by
\begin{equation}
\label{eq:areaRules}
S(E)\!=\!\begin{cases}
2\pi(p_+-p_{-})\!-\!2\tilde S(E), & p_{\pm}\ {\rm on }\ U_{-},\\
2\pi(\frac{1}{2}-p_-)\!-\!2\tilde S(E), & p_{-}\ {\rm on }\ U_{-},\,p_{+}\ {\rm on }\ U_{+},\\
2\pi(\frac{1}{2}+p_+)\!-\!2\tilde S(E), & p_{-}\ {\rm on }\ U_{+},\,p_{+}\ {\rm on }\ U_{-},\\
-2\pi\!+\! 2\tilde S(E),& p_{\pm}\mbox{ on }U_{+}.
\end{cases}
\end{equation}

\begin{figure}[b]
\begin{center}
\includegraphics[width=70mm]{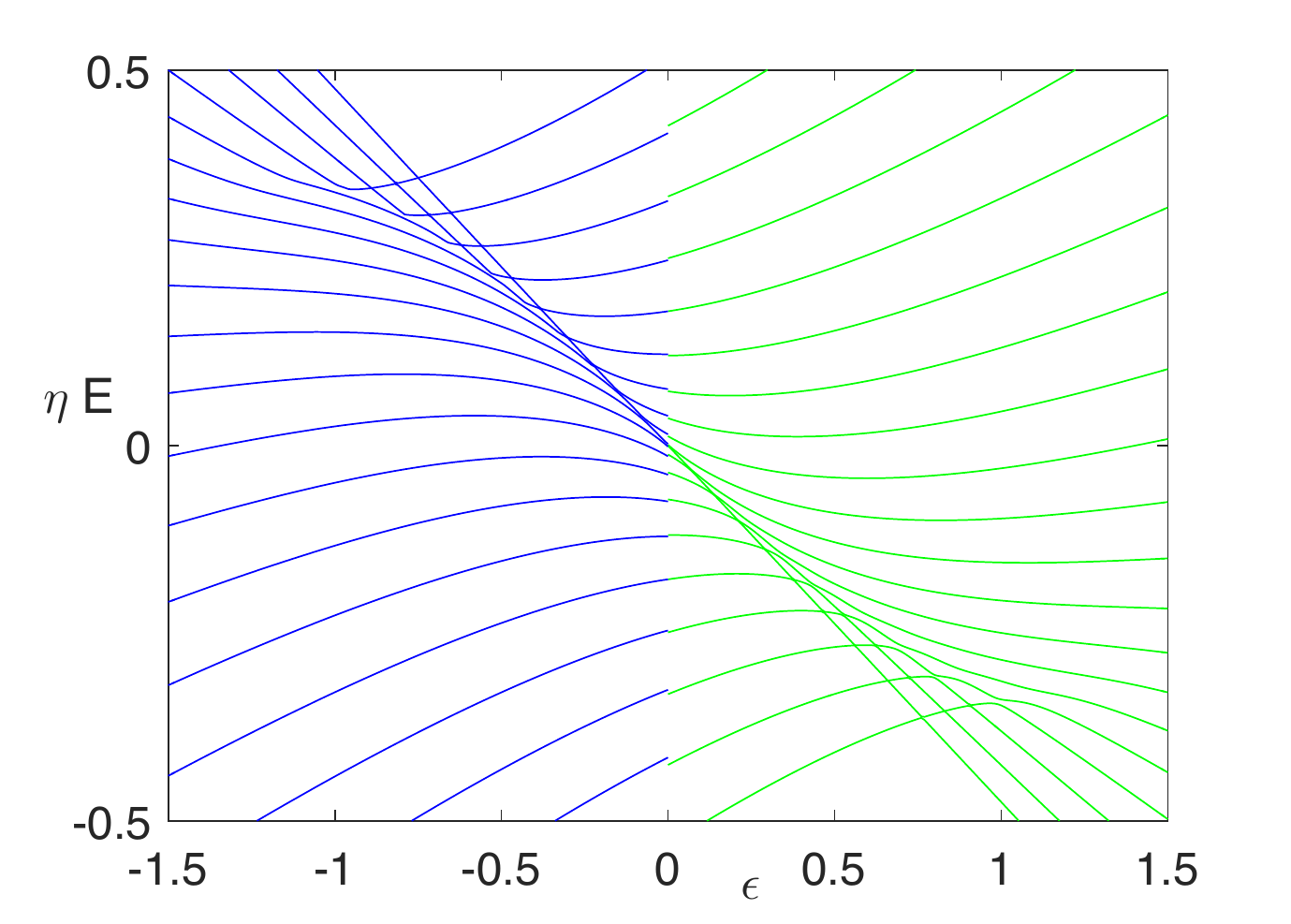}
\includegraphics[width=70mm]{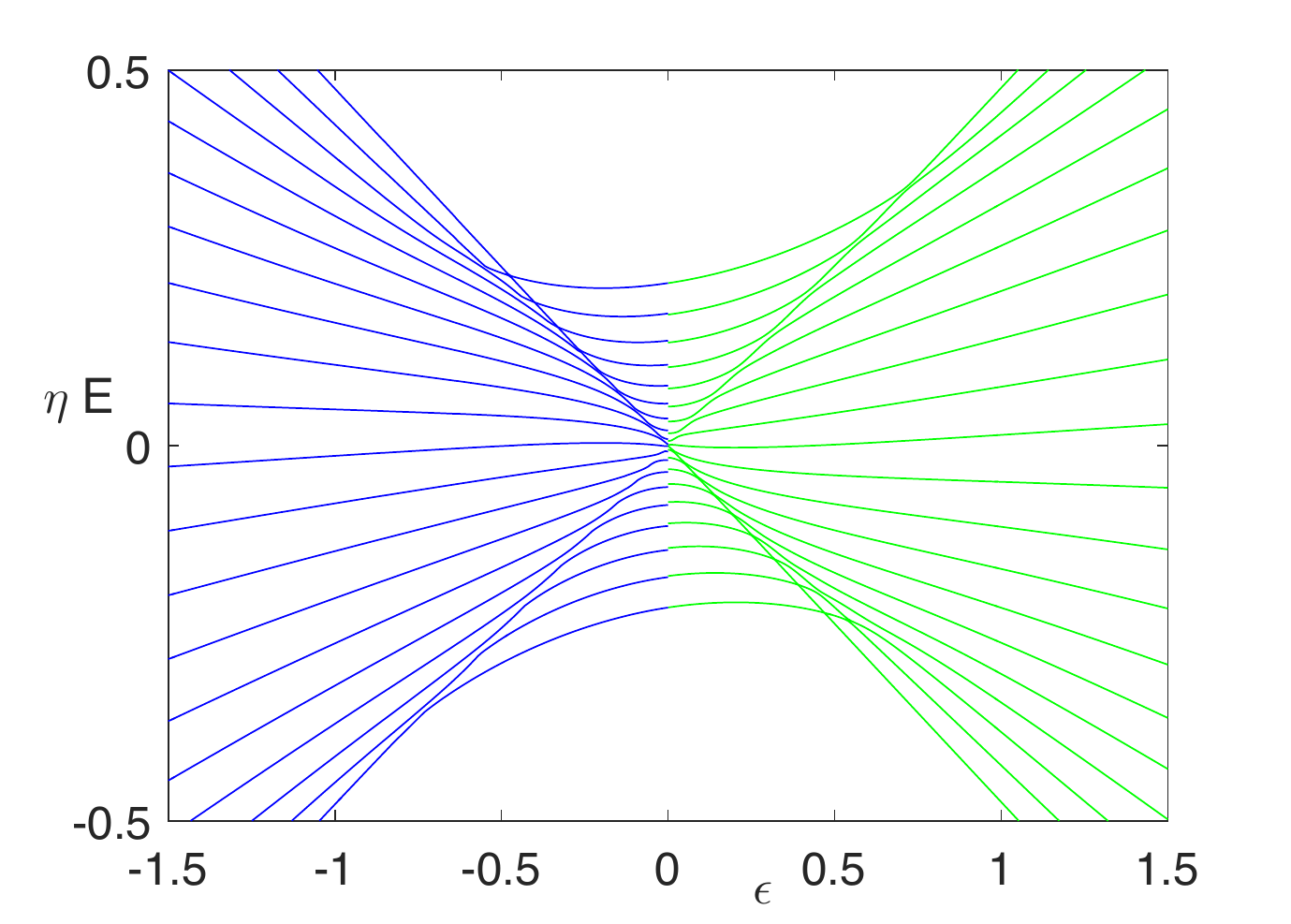}\\
\end{center}
\caption{\label{fig:quantFig2} (Color online) Semiclassical energies $\eta E$  (blue lines, depicted for $\epsilon< 0$) compared with exact
 energies (green lines, depicted for $\epsilon> 0$) for $(m,n) = (4,1)$ (top) and $(m,n) = (4,3)$ (bottom).}
\end{figure}

When $m$ or $n$ are larger than two, there exist parameter values $\epsilon$ and $v$ for which there are two classically allowed regions, i.e., four real turning points in the range $[-\frac{1}{2}, \frac{1}{2}]$, for some energy values. An example of this is shown for the case $(m,m)=(4,1)$ in Figure \ref{fig:quantFig1}. Here we need to take the influence of the potential barrier into account. If one of the minima of $U_-(p)$ is lower, for energies below the upper minimum we can apply the single well quantization condition \eqref{eq:quantisationCondition}. For higher energies, where we have four real turning points $p_-^{(l)}\leq p_+^{(l)}\leq p_-^{(r)}\leq p_+^{(r)}$, we use a WKB matching condition to take into account tunnelling corrections from the classically forbidden barrier, leading to the quantization condition \cite{Froe72,Chil91,07semiMP}
\begin{equation}
\sqrt{1+\kappa^2}\,\cos \left(\tfrac{S_l+S_{r}}{2\eta}-S_\phi\right)=-\cos \left(\tfrac{S_l-S_{r}}{2\eta}\right),\label{sem-quant2}
\end{equation}
where $S_l$ and $S_{r}$ are the phase space areas in the left and right regions, respectively. The term
\begin{equation}
\kappa =\re^{-\pi S_\epsilon}\ ,\
S_\epsilon=\frac{1}{\pi\eta}\int_{p_+^{(l)}}^{p_-^{(r)}}|q(p,E)|\,\rd p
\end{equation}
accounts for tunneling through the barrier, and
\begin{equation}
S_\phi=\arg \Gamma({\textstyle \frac12}+\ri S_\epsilon)-S_\epsilon\,\log |S_\epsilon| +S_\epsilon
\end{equation}
is a phase correction. 

Above the barrier, the inner turning points $p_+^{(l)}$, $p_-^{(r)}$ turn into
a complex conjugate pair and different continuations of the semiclassical quantization have been
suggested  \cite{Chil74,Froe72,Chil91}. Following \cite{Froe72} we use the complex turning points in the formulas for $S_{l,r}$. We modify
the tunneling integral $S_{\epsilon}$ as
\begin{equation}
S_\epsilon=\frac{\ri}{\pi\eta}\int_{p_+^{(l)}}^{p_-^{(r)}}q(p,E)\,\rd p,
\end{equation}
such that the quantity $\kappa$ is positive above the barrier. In equation (\ref{sem-quant2}) we take the real parts of the actions $S_{r,l}$ and $S_\phi$. 

Analogous quantization rules can be applied when the upper potential curve has two maxima.

The results of the semiclassical quantization in comparison with the numerically exact many-particle eigenvalues for different values of $m$ and $n$ are shown in figure \ref{fig:quantFig2}. We observe an excellent agreement, including very well reproduced avoided crossings as we vary the parameter $\epsilon$.

Let us finally turn to a discussion of the many-particle density of states, which can be obtained from the mean-field dynamics on the basis of a semiclassical argument \cite{Grae15}. 
The many-particle density of states
$\rho(E)$ at a scaled energy $E$ 
is (approximately) related to the mean-field period $T(E)=\rd S/\rd E$ 
of the orbit by
\begin{eqnarray}
\label{eq-rhoT}
\rho(E)\! \approx \!\tfrac{1}{2\pi}\,T(E)\!\!=\!\!\frac{1}{\pi}\!\!\int_{p_-}^{p_+}
\!\!\!\!\frac{\rd p}{\sqrt{(U_+(p)\!-\!E)(E\!-\!U_-(p))}}\,,
\end{eqnarray}
where $U_\pm$ are the potential functions \eqref{eq-UpmT}, 
and $p_\pm$ are the turning points, the real valued solutions of $U_\pm(p)=E$ falling into the interval $[- \tfrac12,+\tfrac12]$.
The function under the square root in (\ref{eq-rhoT}),
\begin{eqnarray}
\label{eq-root}
(U_+(p)-E)(E-U_-(p))=v^2r^2(p)-(E-\epsilon p)^2,
\end{eqnarray}
is a polynomial of order $m+n$ in $p$ and the integral (\ref{eq-rhoT}) can be evaluated in 
closed form for $(m,n)=(1,1)$ and $(m,n)=(2,1)$ \cite{07semiMP,Grae15}.
The mean-field period $T(E)$ 
given by the integral  (\ref{eq-rhoT}) can be efficiently  evaluated by means of a Gauss-Mehler quadrature.  
If the fixed point is a center, $T(E)/2\pi$ is given by the inverse
frequency     $\omega=\sqrt{\epsilon^2+v^2 f'(p_c)}$  at the center $p_c$
(see (\ref{fix-om})). In the subcritical parameter region, the period $T$ diverges
logarithmically at the saddle point energies, as already observed before
for dynamics on the Bloch sphere $m=n=1$ in the presence of interactions \cite{07semiMP,Grae14}, and for $(m,n)=(2,1)$ \cite{Grae15}.
Therefore the quantum energy eigenvalues accumulate at the all-molecule
configurations in this regime. Such a level bunching at the classical saddle point energy in this limit can be related to a 
quantum phase transition \cite{Sant06,Pere11,Stra14}. 
We shall now demonstrate this behaviour for several examples.

\begin{figure}
\begin{center}
\includegraphics[width=40mm]{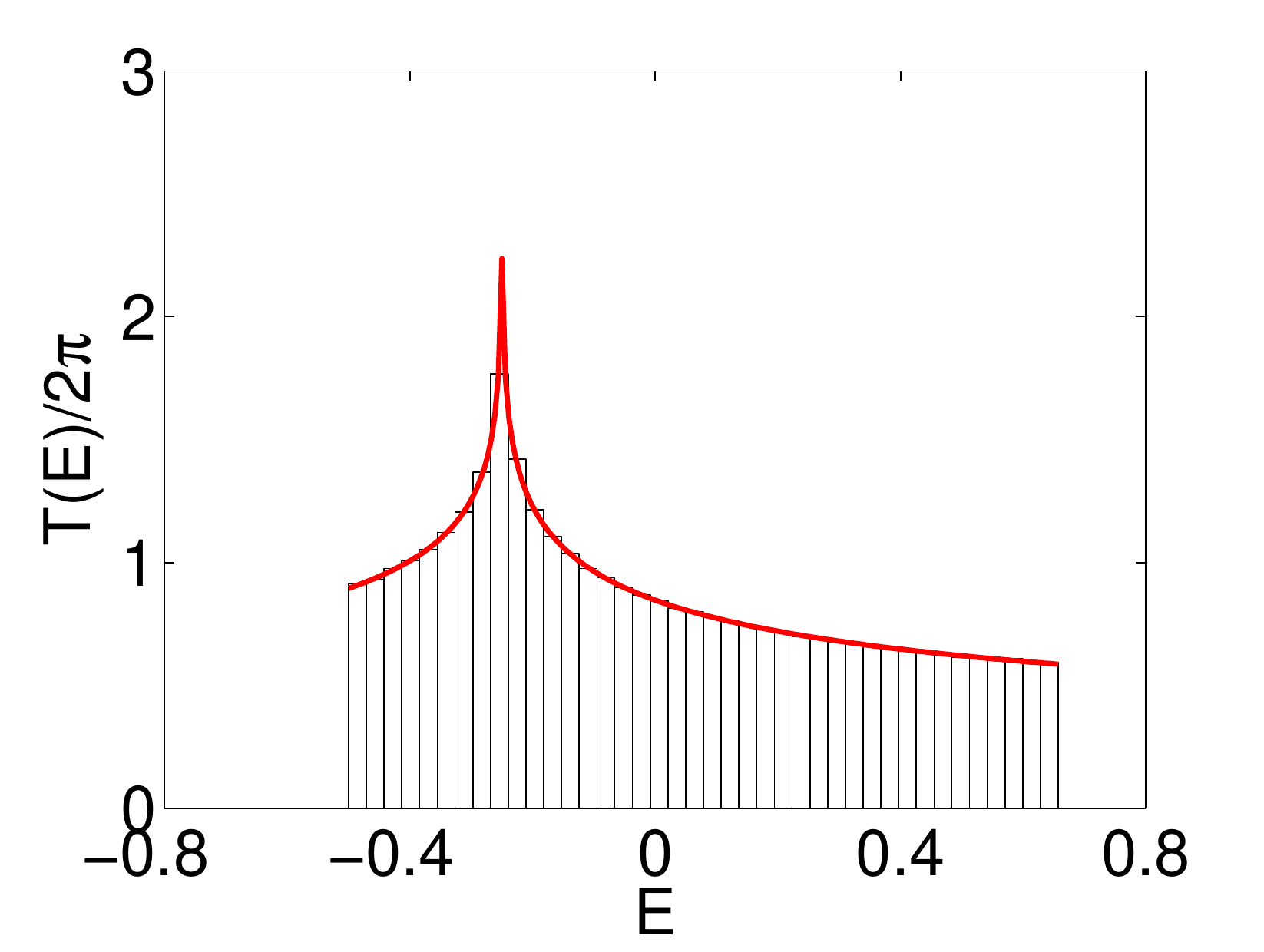}
\includegraphics[width=40mm]{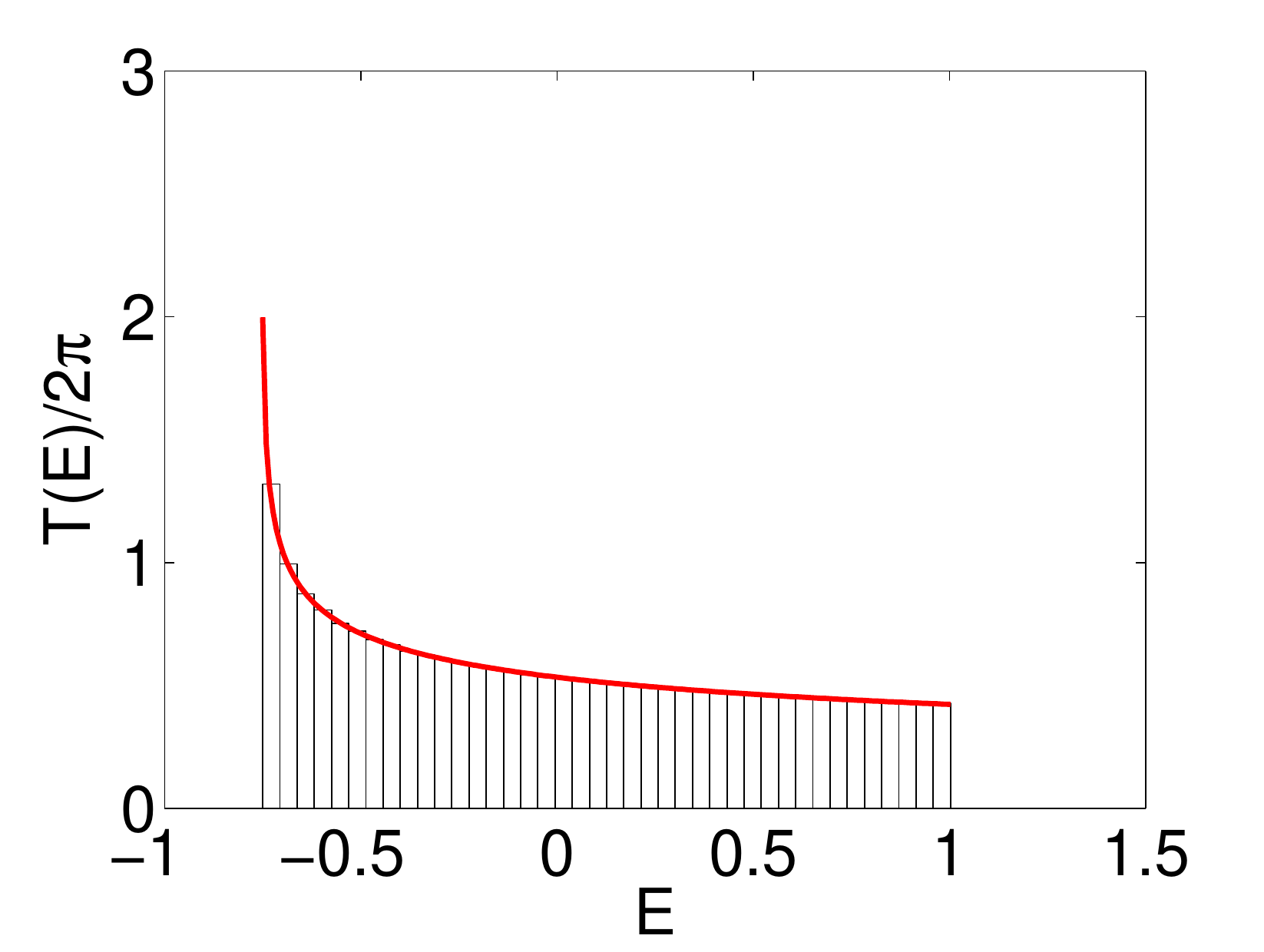}
\end{center}
\caption{\label{fig-den21} (Color online) $(m,n)=(2,1)$\,:\, Mean-field  period $T(E)$ divided by $2\pi$ (red line) and many-particle density of states (histogram) for 
 $N=9000$ particles for $v=1$ and $\epsilon=0.5$ (left) and $\epsilon=1.5$ (right).}
\end{figure}

Figure \ref{fig-den21} shows the mean-field period $T(E)/2\pi$ as well as a
histogram of the many-particle eigenvalues (scaled by a factor $\eta$)
for $N=9000$ particles for $v=1$ in the sub- and supercritical region for the case $(m,n)=(2,1)$.
The density of states is in excellent agreement with the mean-field period. At the boundaries of the
allowed energy interval it is equal to the reciprocal period at the centers. In the subcritical case, there is a divergence at
the energy $-\epsilon/2$ at the south pole in the limit $N\rightarrow \infty$. 

\begin{figure}
\begin{center}
\includegraphics[width=40mm]{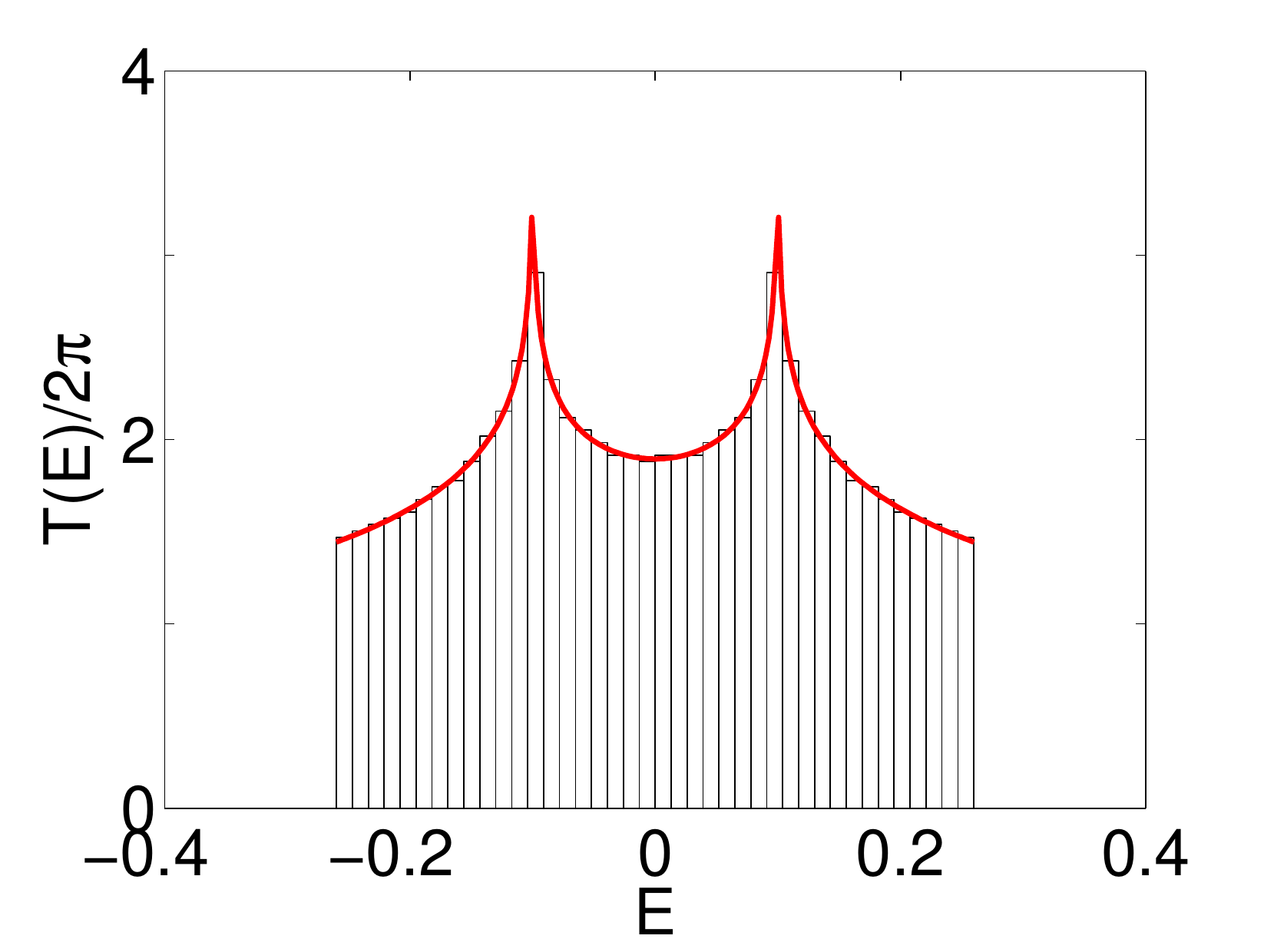}
\includegraphics[width=40mm]{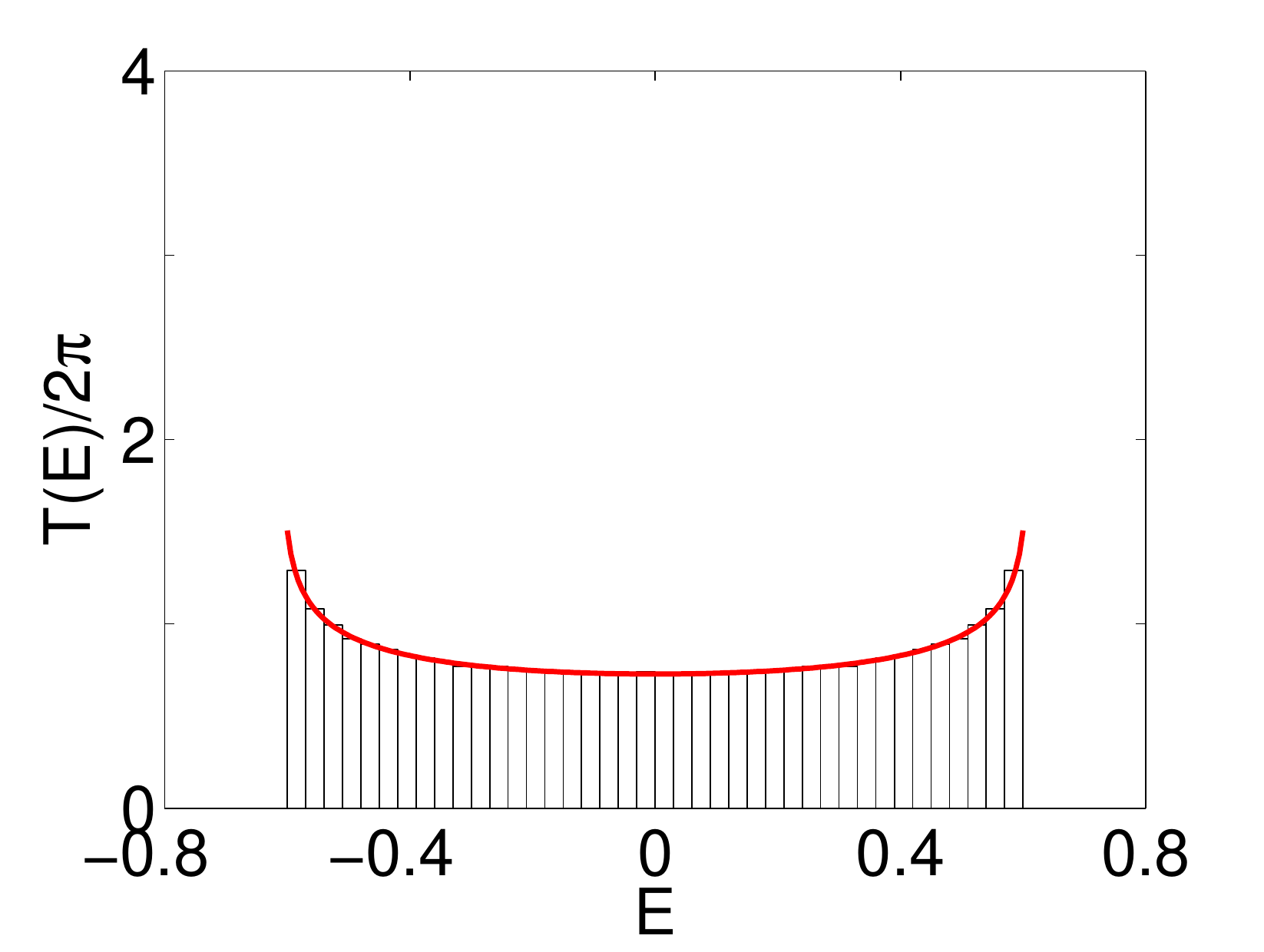}
\end{center}
\caption{\label{fig-den22} (Color online) $(m,n)=(2,2)$\,:\, Mean-field  period $T(E)$ divided by $2\pi$ 
(red line) and many-particle density of states (histogram) for 
 $N=9000$ particles for $v=1$ and $\epsilon=0.2$ (left) and $\epsilon=1.2$ (right).}
\end{figure}

Histograms of the many-particle eigenvalues  for $m=2=n$ for $N=9000$ particles for $v=1$ in comparison with the mean-field periods are shown in figure \ref{fig-den22}, again in the sub- and supercritical regions. Because of $m=n$ the distributions are symmetric. For $\epsilon=0.2$ there are two saddle points at the poles, and hence two singularites,
and for  $\epsilon=1.2$ we only have two centers at the poles with frequency 
$\omega=\sqrt{(v^2-\epsilon^2)/2}$ (see (\ref{fix-om})).\\[2mm]

\begin{figure}
\begin{center}
\includegraphics[width=30mm]{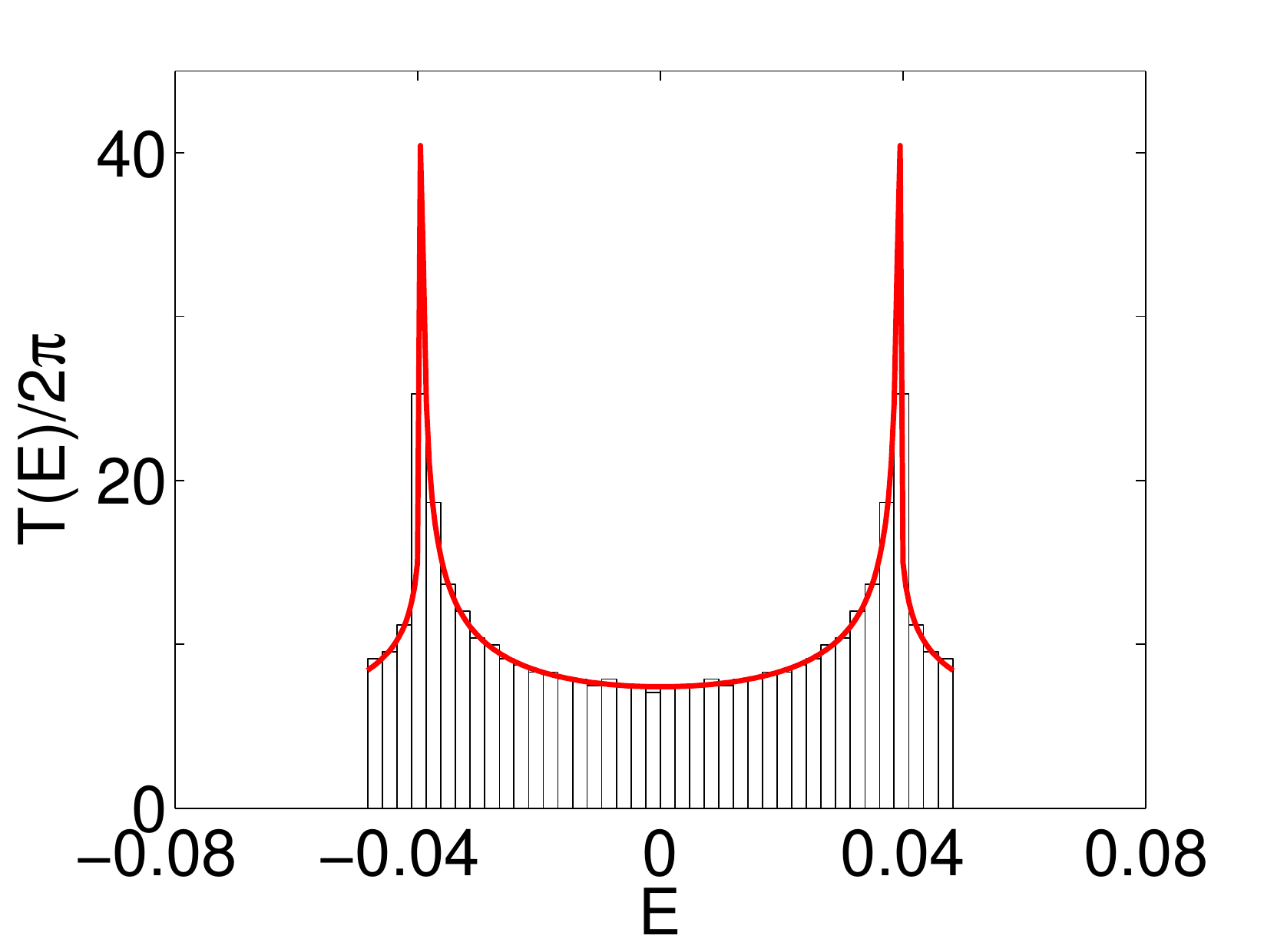}
\hspace*{-4mm}
\includegraphics[width=30mm]{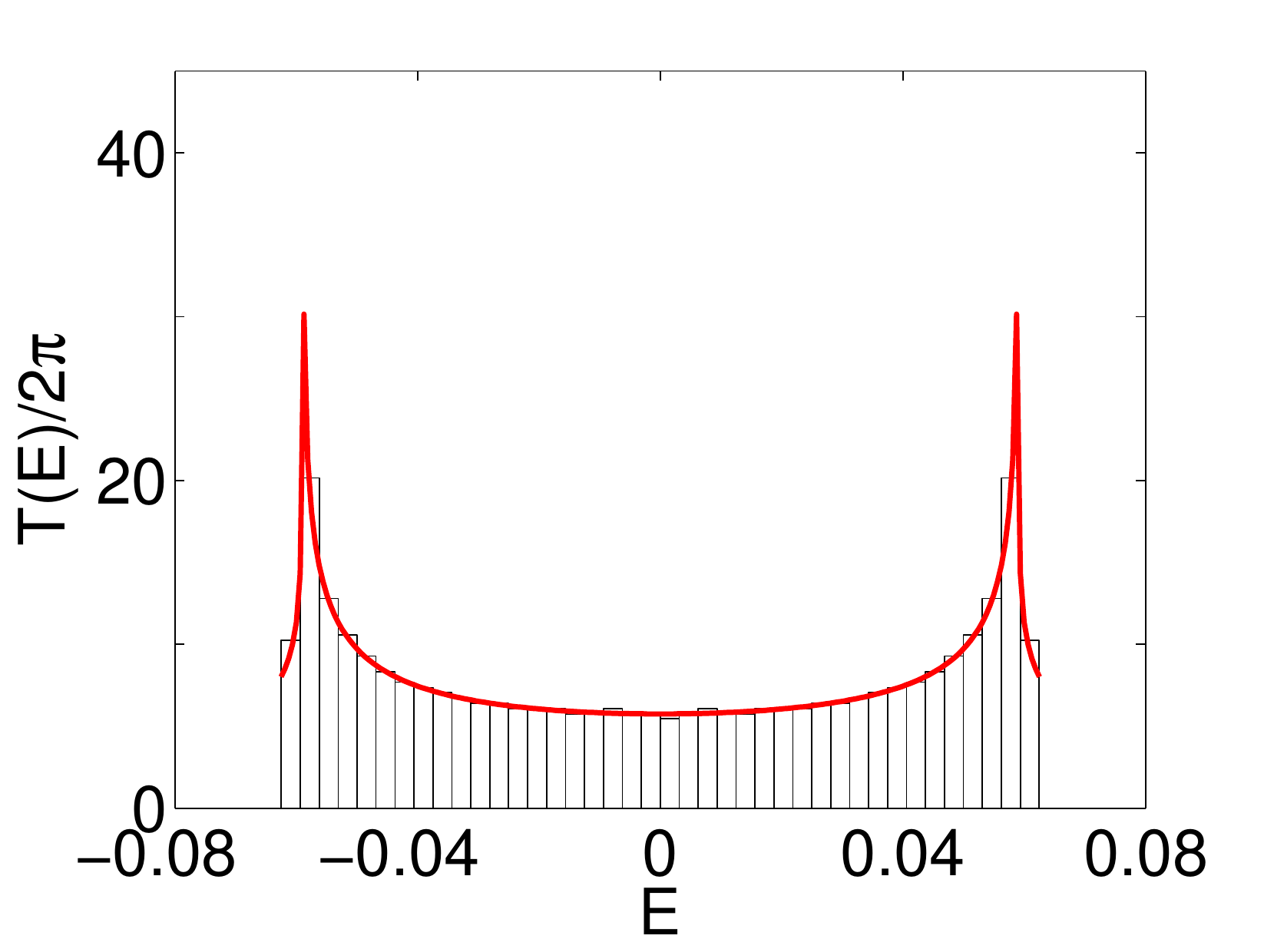}
\hspace*{-4mm}
\includegraphics[width=30mm]{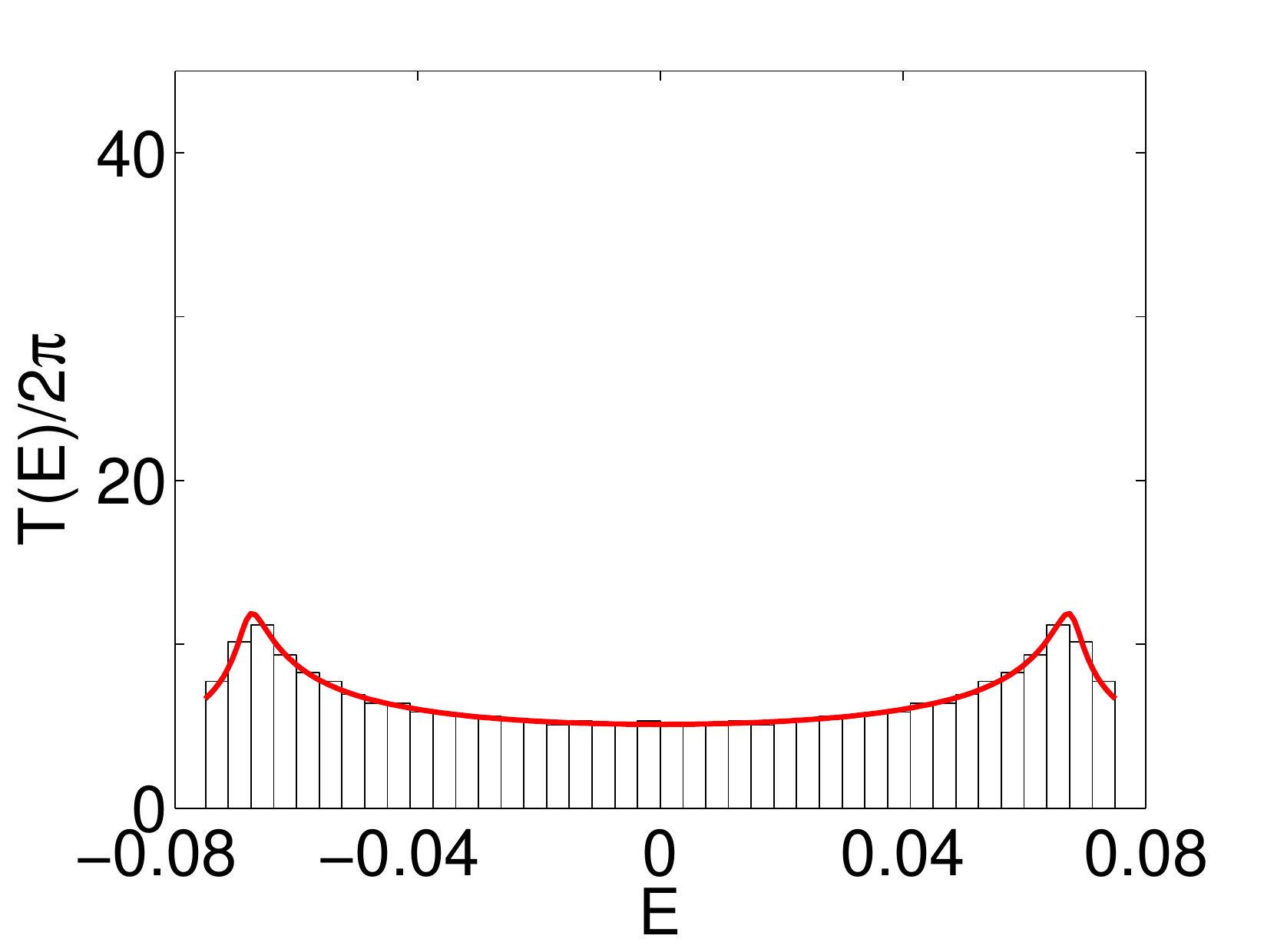}
\includegraphics[width=30mm]{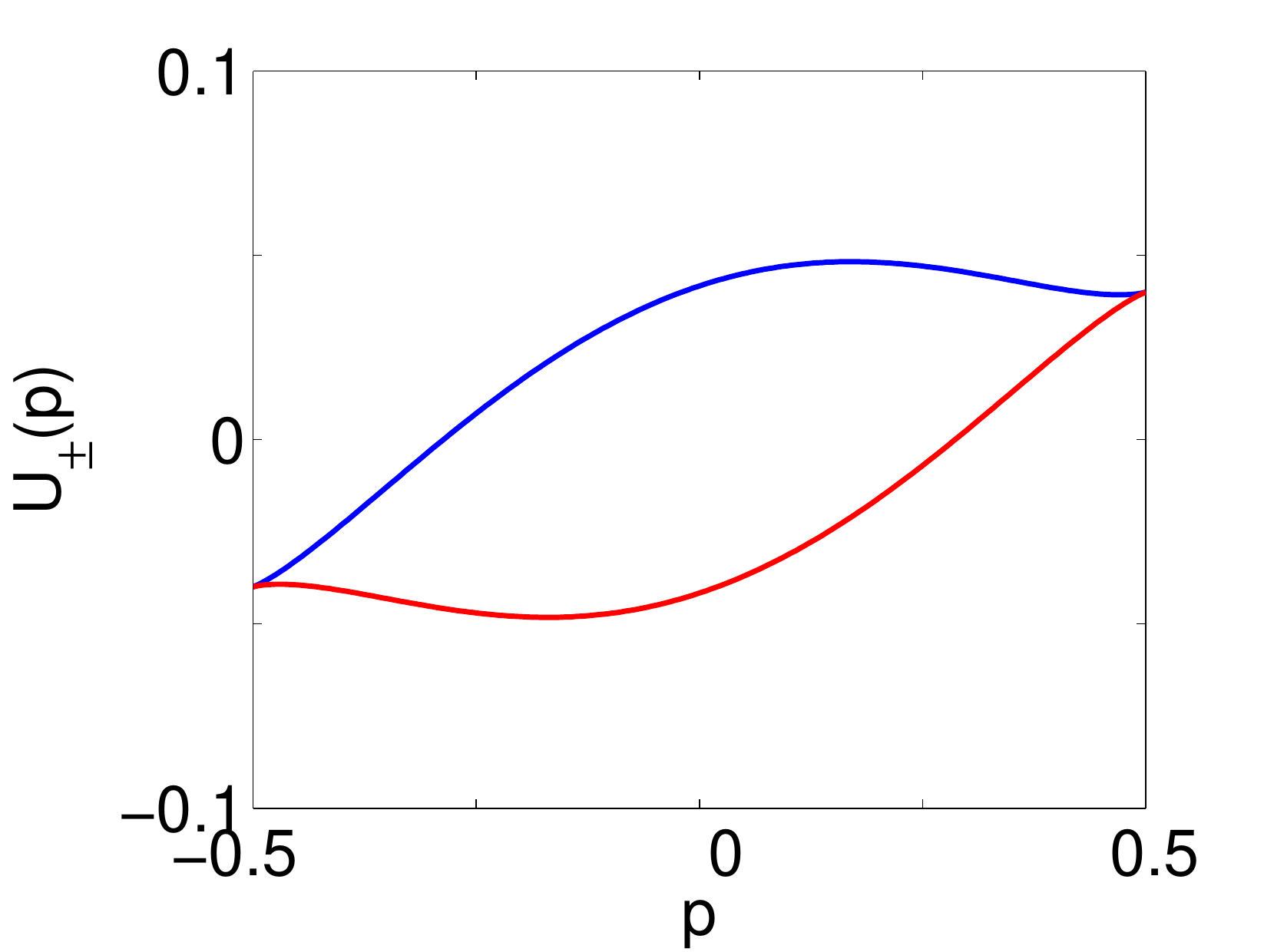}
\hspace*{-4mm}
\includegraphics[width=30mm]{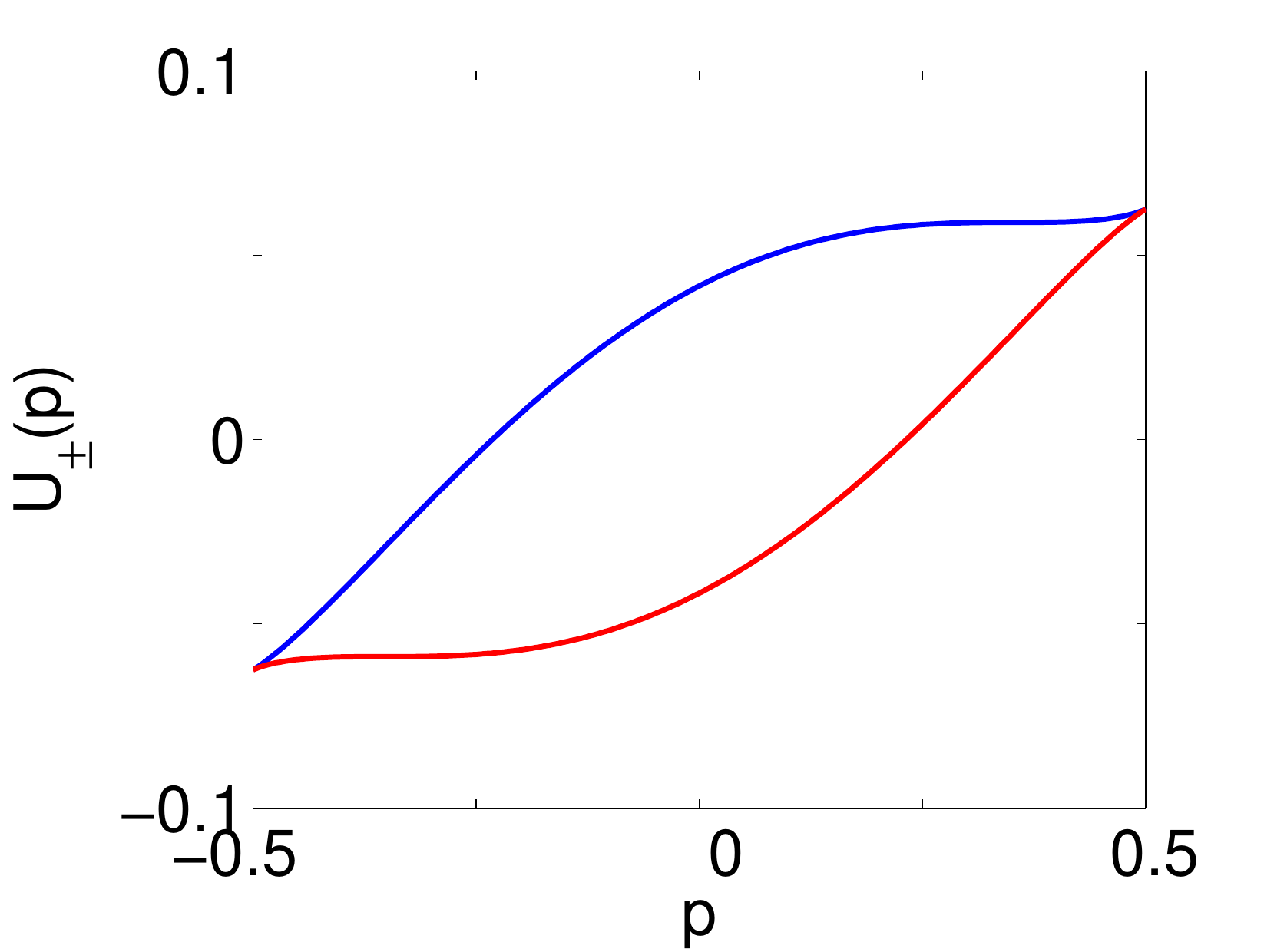}
\hspace*{-4mm}
\includegraphics[width=30mm]{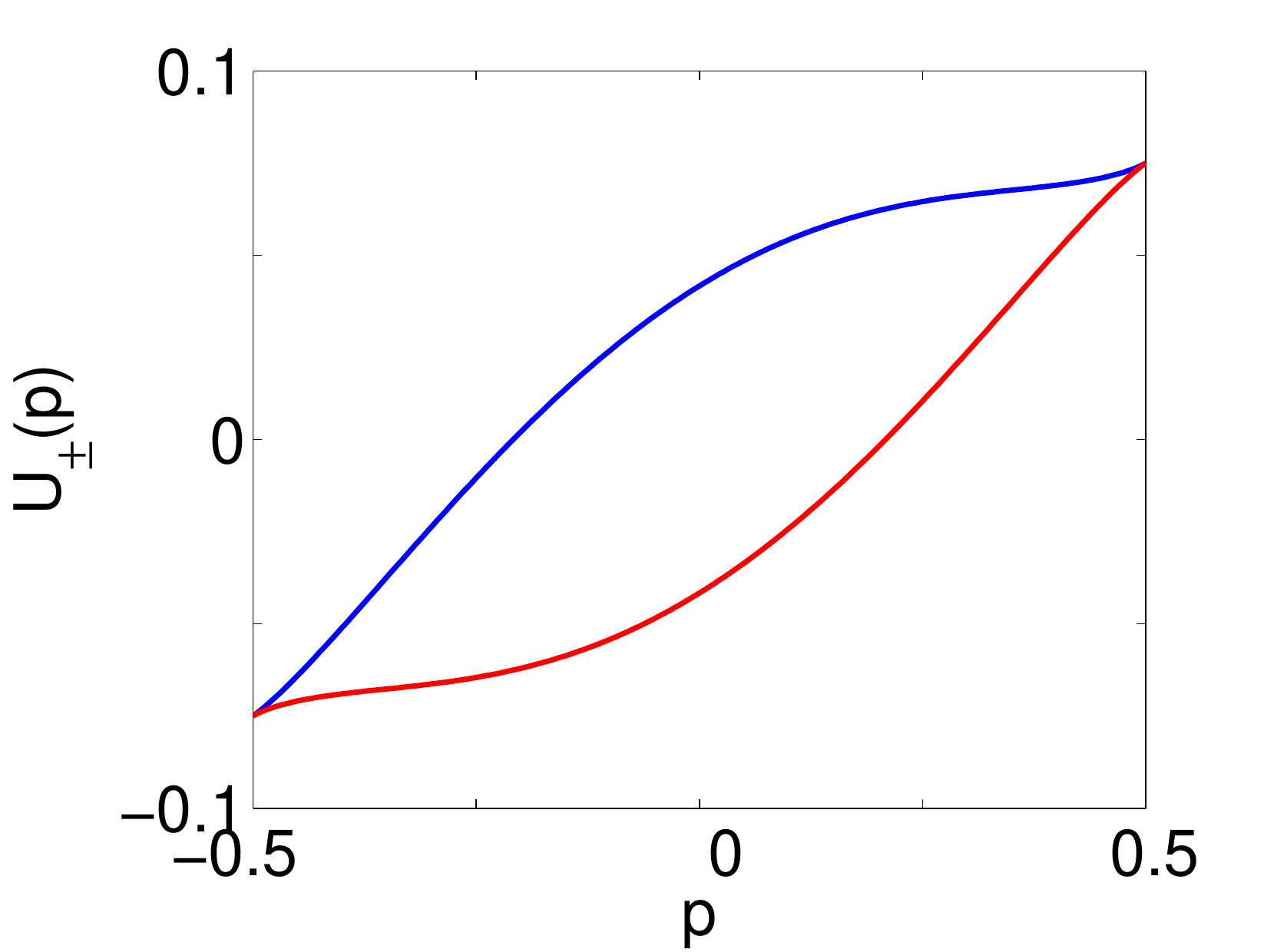}
\end{center}
\caption{\label{fig-den33} (Color online) $(m,n)=(3,3)$\,:\, The top panel shows the mean-field  period $T(E)$ divided by $2\pi$ 
(red line) and many-particle density of states (histogram) for 
 $N=9000$ particles for $v=1$ and $\epsilon=0.08$, $0.125$, $0.15$ (from left to
right). The bottom panel shows the corresponding potential curves $U_+(p)$ (blue) and $U_-(p)$ (red).}
\end{figure}

To understand the densities of states for the case $m=3=n$, depicted in the top row of figure \ref{fig-den33}, it is instructive to have a look at the corresponding potential curves $U_\pm(p)$ defined in (\ref{eq-UpmT}), depicted in the bottom panel of the figure for three values of $\epsilon$. For $\epsilon=0.08$
we are in the subcritical region and the potential $U_-(p)$ has a minimum and a very shallow maximum,
which is hard to identify in the plot, but it must necessarily exist because the slope of both potentials 
$U_\pm(p)$ at $p=-\tfrac12$ is equal to $\epsilon$, i.e.~positive. This shallow maximum
with energy $E_{\rm max-}$ appears as a
fixed point of the dynamics, a saddle point with energy $E_{-}$, and the minimum
with energy $E_{\rm min-}$ as a center. The same is true, of course, for
the potential curve $U_+(p)$ with a saddle energy $E_{+}$, a maximum 
$E_{\rm max+}$ and a minimum
$E_{\rm min+}$.  At the critical value $\epsilon_c=v/8$ the minimum and the
maximum coalesce and disappear for larger values of $\epsilon$.

In the subcritical region $|\epsilon|<\epsilon_c$ there exist two disconnected allowed
potential regions in the energy intervals
$E_{-}<E<E_{\rm max-}$ and $E_{\rm min+}<E<E_{+}$, both contributing to the mean-field density of 
states (\ref{eq-rhoT}), which
therefore shows four steps at the energies of the minima and maxima and
two logarithmic singularities at the energies of the saddle points. 
The top panel in figure \ref{fig-den33} shows histograms of the state density and mean-field periods
for $v=1$ and selected
values of $\epsilon$ in different regions. The case $\epsilon=0.08$ is in the
subcritical region discussed above, for the critical value $\epsilon=0.125$ 
the minima and maxima coincide with the saddle point, shown as 
two singularities of the mean-field period, which disappear in the supercritical regime.
Here, however, they are still observable as peaks in the vicinity of the former singularities,
as shown in the figure for $\epsilon=0.15$. With increasing $\epsilon$ these maxima
decrease. 
\begin{figure}
\begin{center}
\includegraphics[width=60mm]{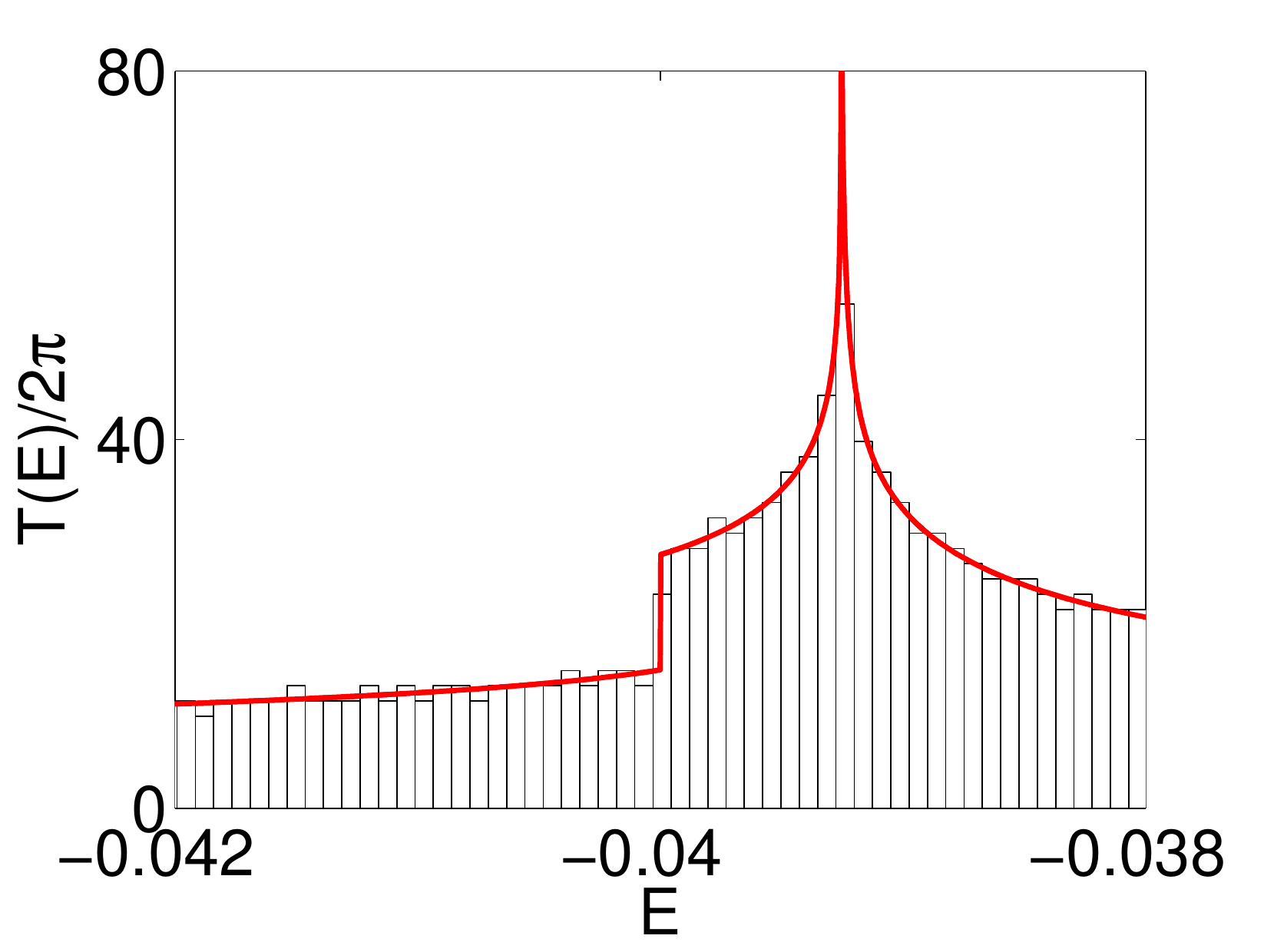}
\end{center}
\caption{\label{fig-den33-mag} (Color online) $(m,n)=(3,3)$\,:\, 
Magnification of the vicinity of the left singularity in figure \ref{fig-den33} 
for  $\epsilon=0.08$, however for $N=72000$ particles.}
\end{figure}
Let us finally explore the subcritical case $\epsilon=0.08$ in more detail to resolve the
structure of the state densities in this regime. A magnification of the neighborhood
of the singularity in figure \ref{fig-den33} (left panel) is shown in figure 
\ref{fig-den33-mag}, however for $N=72000$ particles, where we clearly observe the
step at $E_-=0.04$ in addition to the singularity at the saddle point energy
$E_{\rm max-}$ in the quantum density of states.\\[2mm]

\begin{figure}
\begin{center}
\includegraphics[width=30mm]{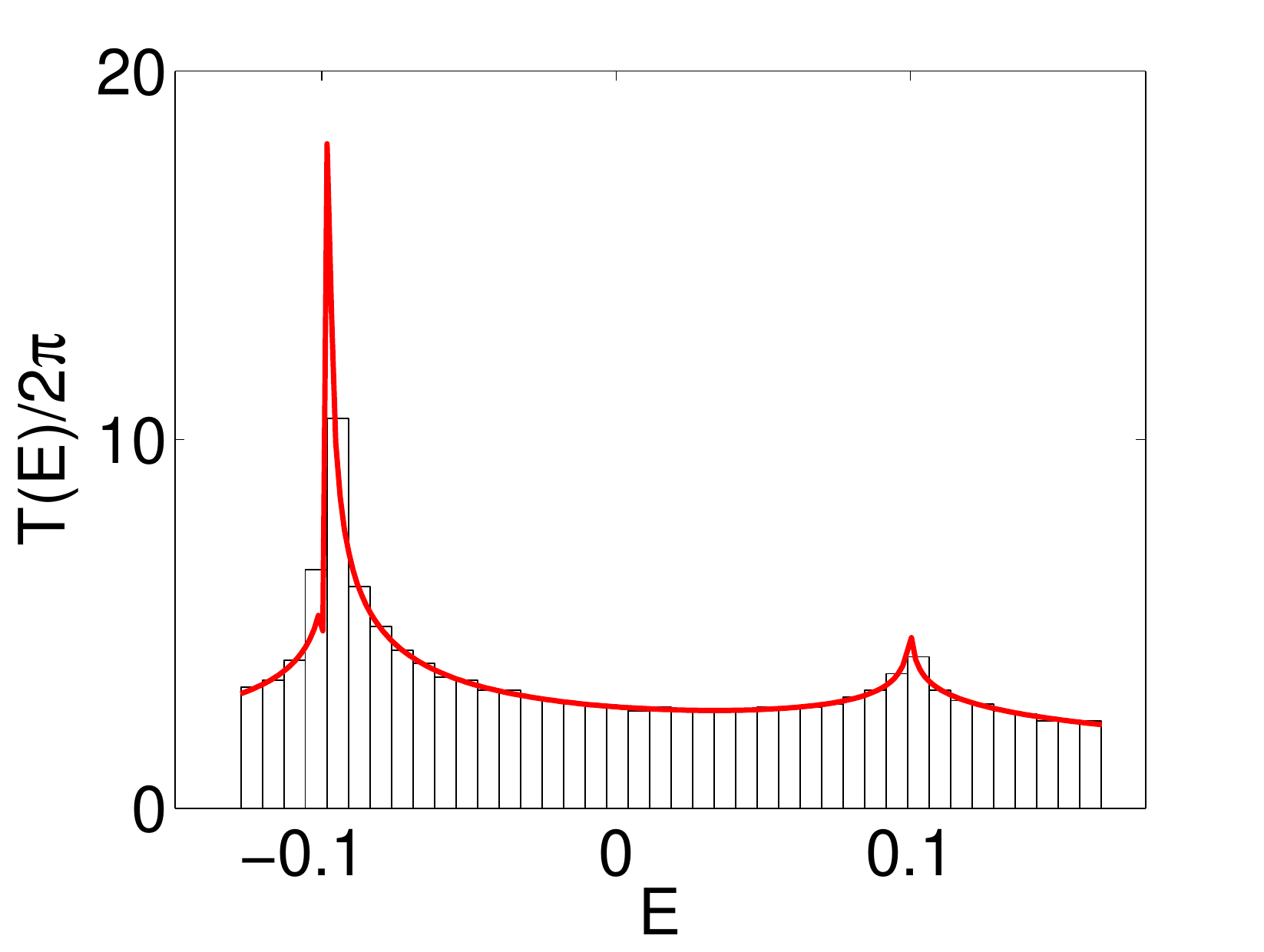}
\hspace*{-4mm}
\includegraphics[width=30mm]{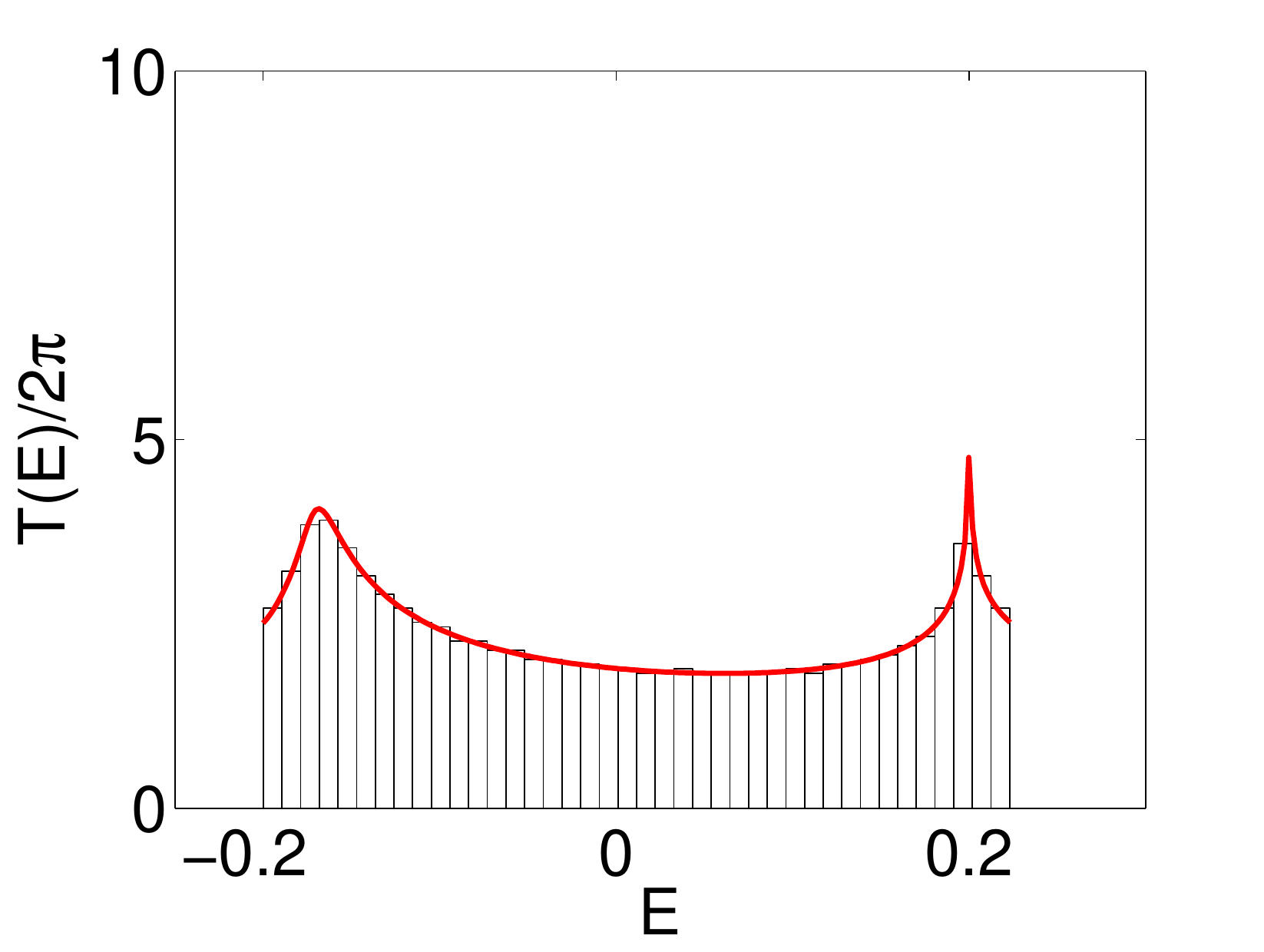}
\hspace*{-4mm}
\includegraphics[width=30mm]{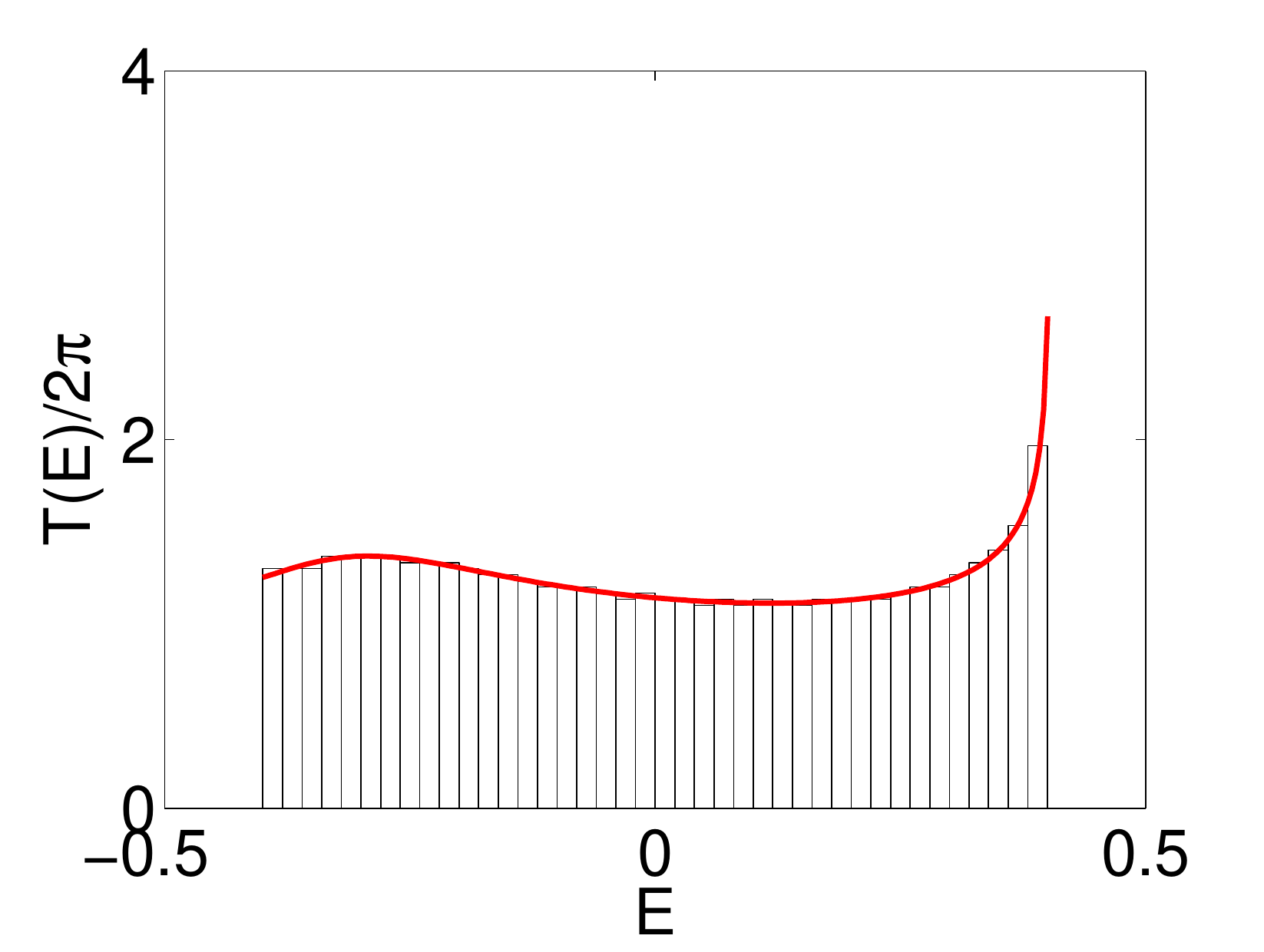}
\end{center}
\caption{\label{fig-den32} (Color online) $(m,n)=(3,2)$\,:\, Mean-field  period $T(E)$ divided by $2\pi$ (red line)
and many-particle density of states (histogram) for 
 $N=9000$ particles for $v=1$ and $\epsilon=0.2,\,0.4,\,0.8$ from left to right.}
\end{figure}

Figure \ref{fig-den32} shows the many-particle densities for $(m,n)=(3,2)$ along
with the mean-field periods in different parameter regions ($\epsilon=0.2$, $0.4$ 
and $0.8$). One of the two singularities for $\epsilon=0.2$ changes
into a maximum for $\epsilon=0.4$ and for $\epsilon=0.8$ also the second
singularity disappeared. Structures, such as the ones observed here can be connected 
to higher order phase transitions \cite{Stra14}.

\section{Summary and Outlook}
The mean-field approximation is quite often indispensable in studies
of multi-particle quantum systems. In addition, as demonstrated above for
simple types of particle conversion systems, it
also offers illuminating tools for understanding the characteristic features
of quantum systems.  The energy spectra, for example, are clearly supported by
the skeleton of mean-field fixed points, showing up, e.g., as boundaries, steps of
singularities of the quantum state densities in the thermodynamic limit of 
large particle numbers. We have further demonstrated that the many-particle 
energies can be accurately recovered from the mean-field description via semiclassical quantization formulas. 

In addition, the present analysis is based on polynomially deformed algebras, where
an interesting connection between quantum (commutator) algebras and classical
(Poisson bracket) ones appeared. The observed differences deserve further studies.
It should also be noted that the transition from quantum to mean-field, the `classicalization',
employed here is quite heuristic and deserves a more sophisticated treatment,
for example in terms of coherent states for deformed algebras as already pointed out
in \cite{Grae15}.

Finally, the present study concentrated on the spectral features of the conversion systems.
A comparison of quantum and mean-field dynamics will also be of interest and corresponding
investigations based on semiclassical phase space densities \cite{07phase} is a topic for future investigations.
\appendix
\section{Polynomial deformations of $\mathbf{su(2)}$}
\label{s-poly-alg}
The deformed algebra is generated by the three elements $\joh=\jod$, $\jp=\jmd$
satisfying 
\begin{eqnarray}
\label{alg-com-1}
[\joh,\jpm]=\pm \jpm\ ,\ [\jp,\jm]=2\hat F(\joh)\,,
\end{eqnarray}
where $[\,.\,,\,.\,]$ is the commutator, and
$\hat F(\joh)=\sum_{j=0}^k\alpha_j\joh^j$ is a polynomial of order $k$.
For $\hat F(\joh)=\joh$ we have $ [\jp,\jm]=2\joh$, i.e.~the Lie algebra $su(2)$,
so that we have a polynomial deformation of  $su(2)$. 
Similar to the Schwinger representation of $su(2)$ the deformed $su(2)$ algebras can be represented via two-mode bosonic creation and annihilation operators \cite{Roce90,Lee10} according to equations (\ref{kxyz}). Using the well known properties of the oscillator algebra one obtains the commutator
\begin{equation}
\label{an-adn-comm}
\big[\ah^m,\am\big]=\Pi_{\mu=1}^{m}(\ad \ah+\mu)-\Pi_{\mu=1}^{m}(\ad \ah+1-\mu),
\end{equation}
which is a polynomial of the number operator $\ad \ah$ whose leading order
term is 
\begin{equation}
\label{an-adn-comm-lo}
\big[\ah^m,\am\big]=m^2(\ad \ah)^{m-1}+\ldots\,.
\end{equation}

It can be shown that the Casimir operator of the deformed $su(2)$ algebra is
 given by
\begin{equation}
\label{casi-1}
\hat C = \jm\jp+\hat \phi(\joh)
\end{equation}
where $\hat \phi(\joh)$ is a polynomial in $\joh$ of order $k+1$ with $\hat \phi(0)=0$, which can be
expressed in terms of Bernoulli polynomials $B_n(z)$ and Bernoulli numbers $B_n=B_n(0)$ as
\begin{equation}
\label{phi-1}
\hat \phi(\joh)=2\sum_{j=0}^k\frac{(-1)^{j+1}}{j+1}\,\alpha_j\,\big(B_{j+1}(-\joh)-B_{j+1}\big)\,,
\end{equation}
related to $\hat F(\joh)$ by
\begin{equation}
\label{F-phi-1}
\hat F(\joh)=\tfrac{1}{2}\big(\hat \phi(\joh)-\hat \phi(\joh-1)\big).
\end{equation}
 (see, e.g., \cite{Delb93} and references therein). Up to third order, $k=3$, (\ref{phi-1}) yields
 \begin{eqnarray}
&&\hat \phi(\joh)=\Big(2\alpha_0+\alpha_1+\tfrac{\alpha_2}{3}\Big)\joh
+\Big(\alpha_1+\alpha_2+\tfrac{\alpha_3}{2}\Big)\joh^2\nn\\
&&\qquad\qquad\qquad \qquad+\Big(\tfrac{2\alpha_2}{3}+\alpha_3\Big)\joh^3
+\tfrac{\alpha_3}{2}\joh^4 \label{phi-k=3}
\end{eqnarray}
(see also \cite{Debe00,Beck00}). Alternatively, with
\begin{equation}
\label{jxy}
\jx=\tfrac{1}{2}\big(\jp +\jm\big)\  ,\ \jy=\tfrac{1}{2\ri}\big(\jp-\jm\big)\ ,\ \jz=\joh\,,
\end{equation}
and 
\begin{eqnarray}
\label{alg-com-2}
[\jy,\jz]=\ri\jx\ ,\ [\jz,\jx]=\ri \jy\ ,\ [\jx,\jy]=\ri\,\hat F(\jz)
\end{eqnarray}
the Casimir (\ref{casi-1}) is written as
\begin{eqnarray}
\label{casi-2}
\hat  C&=&\jx^2\!+\!\jy^2\!-\!\hat F(\jz)\!+\!\hat \phi(\jz)\nonumber\\
&=&\jx^2\!+\!\jy^2\!+\!\tfrac{1}{2}\big(\hat \phi(\jz)\!+\!\hat \phi(\jz\!-\!1)\big).
\end{eqnarray}
For polynomials up to third order (\ref{phi-k=3}) implies
\begin{eqnarray}
&&\hat C =\jx^2+\jy^2-\alpha_0+
\Big(2\alpha_0+\tfrac{\alpha_2}{3}\Big)\jz\nn\\
&&\qquad\quad +\Big(\alpha_1+\tfrac{\alpha_3}{2}\Big)\jz^2
+\tfrac{2\alpha_2}{3}\,\jz^3
+\tfrac{\alpha_3}{2}\,\jz^4\,.\label{casi-3}
\end{eqnarray}
For the linear case $\hat F(\jz)=\jz$ we have $\hat \phi(\jz)=\jz+\jz^2$ and
$\hat C= \jx^2+\jy^2+\jz^2$.\\[1mm]

The commutator and the Casimir operator can alternatively be expressed using an 
auxiliary polynomial of order $m+n$ in $\tfrac{\hn}{2mn}$ and $\kz$ defined as\begin{eqnarray}
\label{APmn}
\hat P(\jz)\!=\!
\prod_{\mu=1}^m\!(\tfrac{\hn}{2mn}\!+\!\jz\!+\!\tfrac{\mu}{m})
\prod_{\nu=1}^n\!(\tfrac{\hn}{2mn}\!-\!\jz\!-\! 1\!+\!\tfrac{\nu}{n}),
\end{eqnarray}
with
\begin{eqnarray}
\label{Pmn-km1}
\hat P(\jz) \longleftrightarrow \hat P(-\jz -1)\ \ \textrm{for}\ \ 
(m,n) \longleftrightarrow  (n,m).
\end{eqnarray}
We then define the operator functions
\begin{eqnarray}
\hat F(\jz) &=&-\frac{n^n m^m}{2N^{m+n-2}}\,\big(\hat P(\jz)-\hat
P(\jz-1)\big)\,,\label{Afop}\\[1mm]
\hat G(\jz) &=&-\frac{n^n m^m}{2N^{m+n-2}}\,\big(\hat P(\jz)+\hat
P(\jz-1)\big).\label{Agop}
\end{eqnarray}
From (\ref{Pmn-km1}) we find
\begin{eqnarray}
\label{fgsym1}
\hat F(\jz) \longleftrightarrow -\hat F(-\jz)\  ,\ 
\hat G(\jz) \longleftrightarrow \hat G(-\jz)
\end{eqnarray}
for $(m,n) \longleftrightarrow  (n,m)$ 
and therefore for  $m=n$ the symmetries
\begin{eqnarray}
\label{fgsym2}
\hat F(-\jz)=-\hat F(\jz) \quad \textrm{and}\quad  
\hat G(-\jz)=\hat G(\jz)\,,
\end{eqnarray}
i.e.~$\hat F(\jz)$ and $\hat G(\jz)$ are odd or even polynomials. 
The leading order term of the polynomials
$\hat P(\jz)$ and $\hat P(\jz-1)$ is equal to $(-1)^{n+1}\jz^{(m+n)}$ and hence 
$\hat G(\jz)$ or $\hat F(\jz)$ are polynomials in $\jz$ of order  $m+n$ or $m+n-1$, respectively. The function $\hat F$ is the one appearing in the commutator (\ref{alg-com-1}) and the Casimir operator can be written as 
\begin{eqnarray}
\hat C=\jx^2+\jy^2+\hat G(\jz).
\end{eqnarray}
It should be noted that the relations above depend on the Lie bracket of the
algebra. They are derived for the commutator bracket and are different
for the Poisson bracket (see the footnote in \cite{Roce90}). In both cases we find
a relation between the polynomial $\hat F(\hat J_z)$ or $f(s_z)$ appearing
in the polynomial extension of the Lie brackets and the polynomial 
$\hat G(\hat J_z)$ or $g(s_z)$ appearing in the Casimir operator. This relation is simple 
in the classical algebra, namely $g'(s_z)=2f(s_z)$ (see equation  (\ref{g-f-cl}),
and more elaborate in the quantum algebra.

Let us finally evaluate the leading terms in the limit of large $N$. 
With the abbreviations $\hat A_\pm=\tfrac{\hn}{2mn}\pm \jz$ one obtains from 
(\ref{APmn}) and (\ref{Afop}), (\ref{Agop})
\begin{eqnarray}
\label{Pasy1}
\hat P(\jz)&=&\hat A_+^{m}\hat A_-^{n}-\tfrac{n-1}{2}\,\hat A_+^{m}\hat A_-^{n-1}\nn\\
&&
+\tfrac{m+1}{2}\,\hat A_+^{m\!-\!1}\hat A_-^{n}+\ldots
\end{eqnarray}
\begin{eqnarray}
\label{Pasy2}
\hat P(\jz-1)&=&\hat A_+^{m}\hat A_-^{n}+\tfrac{n+1}{2}\hat A_+^{m}\hat A_-^{n-1}\nn\\
&&
-\tfrac{m\!-\!1}{2}\hat A_+^{m\!-\!1}\!\hat A_-^{n}+\ldots
\end{eqnarray}
and
\begin{eqnarray}
\label{FGasy}
\hat F(\jz)\!\!\!&=&\!\!\!\frac{n^n m^m(
n\,\hat A_+^{m}\hat A_-^{n-1}\!\!-\!m\hat A_+^{m-1}\hat A_-^{n}\!\!+\!\ldots)}{2N^{m+n-2}}\\[1mm]
\hat G(\jz)\!\!\!&=&\!\!\! -\frac{n^n m^m(\hat A_+^{m}\hat A_-^{n}+\ldots)}{2N^{m+n-2}}.
\end{eqnarray}
\begin{acknowledgments}
The authors thank Kevin Rapedius for valuable comments on the manuscript. E.M.G. acknowledges support from the Royal Society via a University Research Fellowship (Grant. No. UF130339). A.R. acknowledges support from the Engineering and Physical Sciences Research Council via the Doctoral Research Allocation Grant No. EP/K502856/1.
\end{acknowledgments}
%

\end{document}